\documentclass[twocolumn,preprintnumbers]{aastex63} 
\usepackage{afterpage}
\usepackage{capt-of}
\usepackage{diagbox}
\usepackage[figuresright]{rotating}
\usepackage{amsmath}
\usepackage{amssymb,amsmath,latexsym,graphics, graphicx,epsfig,multirow,comment,hyperref,appendix,feyn,slashed,xcolor,afterpage, makecell} 
\usepackage{booktabs}
\usepackage{tabularx}
\usepackage{xspace}

\newcommand{\FeH}{\text{[Fe/H]} }

\newcommand {\be} {\begin {equation}}
\newcommand {\ee} {\end {equation}} 

\newcommand {\bes} {\begin {equation*}}
\newcommand {\ees} {\end {equation*}}

\newcolumntype{L}[1]{>{\raggedright\let\newline\\\arraybackslash\hspace{0pt}}m{#1}}
\newcolumntype{C}[1]{>{\centering\let\newline\\\arraybackslash\hspace{0pt}}m{#1}}
\newcolumntype{R}[1]{>{\raggedleft\let\newline\\\arraybackslash\hspace{0pt}}m{#1}}

\usepackage{csvsimple}

\newcommand{\Z}{\mathbb{Z}}
\newcommand{\N}{\mathbb{N}}
\newcommand{\R}{\mathbb{R}}
\newcommand{\C}{\mathbb{C}}

\providecommand{\ie}{\emph{i.e.}\xspace}
\providecommand{\eg}{\emph{e.g.}\xspace}

\providecommand{\Gaia}{\emph{Gaia}\xspace}
\providecommand{\Latte}{\emph{Latte}\xspace}
\providecommand{\FIRE}{\textsc{Fire}\xspace}

\providecommand{\mi}{\texttt{m12i}\xspace}

\renewcommand{\vec}[1]{\mathbf{#1}}

\providecommand{\kpc}{\,\text{kpc}}

\makeatletter
\newcommand\footnoteref[1]{\protected@xdef\@thefnmark{\ref{#1}}\@footnotemark}
\makeatother

\def\CN{\mathcal{N}}

\DeclareRobustCommand{\Sec}[1]{Sec.~\ref{#1}}

\DeclareRobustCommand{\Tab}[1]{Table~\ref{#1}}

\DeclareRobustCommand{\Fig}[1]{Fig.~\ref{#1}}
\DeclareRobustCommand{\Figs}[2]{Figs.~\ref{#1} and \ref{#2}}

\DeclareRobustCommand{\Eq}[1]{Eq.~\eqref{#1}}

\newcommand{\beq}{\begin{equation}}
\newcommand{\eeq}{\end{equation}}

\newcommand {\kms} {\,\,\text{km}/\text{s}}
\newcommand{\V}[1]{{ \bf #1}}

\definecolor{colorTC}{rgb}{.2,.7,.2}

\begin{document}


\title{
\large Chasing Accreted Structures within Gaia DR2 using Deep Learning
}

\author{Lina Necib}
\email{lnecib@caltech.edu}
\affiliation{Walter Burke Institute for Theoretical Physics,
California Institute of Technology, Pasadena, CA 91125, USA}
\affiliation{Department of Physics and Astronomy,
University of California Irvine, Irvine, CA 92697, USA}

\author{\vspace{-0.5cm}Bryan Ostdiek}
\affiliation{Institute of Theoretical Science, Department of Physics, University of Oregon, Eugene, OR 97403, USA}
\email{bostdiek@g.harvard.edu}

\author{\vspace{-0.5cm}Mariangela Lisanti}
\affiliation{Department of Physics, Princeton University, Princeton, NJ 08544, USA}

\author{\vspace{-0.5cm}Timothy Cohen}
\affiliation{Institute of Theoretical Science, Department of Physics, University of Oregon, Eugene, OR 97403, USA}

\author{\vspace{-0.5cm}Marat Freytsis}
\affiliation{Raymond and Beverly Sackler School of Physics and Astronomy, Tel-Aviv University, Tel-Aviv
69978, Israel}
\affiliation{School of Natural Sciences, Institute for Advanced Study, Princeton, NJ 08540, USA}

\author{\vspace{-0.5cm}Shea Garrison-Kimmel}
\affiliation{TAPIR, California Institute of Technology, Pasadena, CA 91125, USA}

\begin{abstract}
 In previous work, we developed a deep neural network classifier that only relies on phase-space information to obtain a catalog of accreted stars based on the second data release of \Gaia (DR2). 
In this paper, we apply two clustering algorithms to identify velocity substructure within this catalog.  We focus on the subset of stars with line-of-sight velocity measurements that fall in the range of Galactocentric radii $r \in [6.5, 9.5]\kpc$ and vertical distances $|z| < 3\kpc$. Known structures such as \Gaia Enceladus and the Helmi stream are identified. The largest previously-unknown structure, Nyx, is a vast stream consisting of at least 90 stars in the region of interest. This study displays the power of the machine learning approach by not only successfully identifying known features, but also discovering new kinematic structures that may shed light on the merger history of the Milky Way. \\

\end{abstract}

\section{Introduction}

The paradigm of hierarchical structure formation describes how galaxies grow in a Lambda Cold Dark Matter universe~\citep{White:1977jf}.  During this process, large galaxies like our Milky Way gain the majority of their mass by capturing and absorbing smaller satellites.  As a satellite galaxy falls onto a host, it is torn apart by violent tidal forces that strip both its dark matter halo and stars.  Debris from this process is left scattered about the host galaxy as a fossil remnant of the accretion events.  Assuming that not enough time has passed for these objects to fully virialize, such remnants manifest as distinctive features in phase space: stellar clumps, streams, and clouds of tidal debris~\citep{Johnston:1996sb, Johnston:1997fv, Bullock:2005pi, Font:2005rm, Robertson:2005gv, 2011MNRAS.416.2802F}.  If such signatures can be identified, one can use their properties to reconstruct aspects of the host galaxy's evolution.

The study of spatial and kinematic substructure in the Milky Way's stellar halo has a long history.  Various works have focused on modeling its large-scale  components, discovering that the stellar halo can be separated into (at least) an inner and outer component based on kinematics  based on distinctions in spatial density profiles, stellar orbits, and spectroscopic metallicities  \citep{Carollo:2007xh, 2010ApJ...712..692C,2012ApJ...746...34B}  photometry (Hess diagrams) in combination with spatial density distributions \citep{2010ApJ...714..663D},  and photometric metallicity estimates in combination with kinematics inferred from proper motions alone \citep{2013ApJ...763...65A, 2015ApJ...813L..28A}.  These works targeted the overall structure of the stellar halo. In parallel, other studies focused on identifying the substructures that built the accreted halo. 
The Sagittarius Stream provides the most stunning example of an ongoing merger~\citep{Johnston:1995vd,Ivezic:2000ua, Yanny:2000ty, Ibata:2000pu}, as it traces several orbits of the infalling Sagittarius dwarf~\citep{1994Natur.370..194I}.  Numerous tidal streams, some of which may be associated with disrupting globular clusters, have also been discovered; see~\citet{2016ASSL..420.....N} for a review.  These include the GD-1~\citep{Grillmair_2006}, Pal-5~\citep{Odenkirchen:2000zx}, and Ophiuchus~\citep{10.1093/mnrasl/slu089} streams.  In the case of older mergers, velocity coherence is still preserved although spatial coherence may be lost~\citep{1999MNRAS.307..495H,Lisanti:2011as,Lisanti:2014dva}.  A plethora of such kinematic substructure has been found throughout the stellar halo, \eg ECHOS~\citep{Schlaufman:2009jv, Schlaufman:2011kf, 2012ApJ...749...77S}.

The advent of data from the \Gaia satellite~\citep{2018arXiv180409365G} has already revolutionized our ability to reconstruct the Milky Way's history, given the unprecedented number of 5D and even 6D stellar phase-space measurements.  To date, several additional stellar streams and clumps have been discovered using \Gaia data~\citep{2018MNRAS.475.1537M, 2018ApJ...860L..11K, 2018MNRAS.477.4063M, 2019A&A...625A...5K, 2019arXiv190403185M, 2019A&A...622L..13M, 2019NatAs.tmp..258I, 2019ApJ...872..152I}. \Gaia Enceladus (also referred to as the \Gaia Sausage) is the largest of these new structures~\citep{2018MNRAS.478..611B,2018Natur.563...85H}, and is believed to be the remnant of a significant merger that occurred at an estimated redshift of $\sim 1$--3~\citep{2018ApJ...863L..28M}.  Today, Enceladus' stellar debris is highly radial and more metal-rich than the rest of the stellar halo; additionally, it constitutes the majority of the accreted stellar fraction in the inner Milky Way~\citep{2018ApJ...862L...1D, 2019ApJ...874....3N, 2018arXiv180704290L,2019MNRAS.487L..47V}.   

The focus of this paper is on searching for kinematic substructures in the subset of \Gaia data that includes radial velocities. In particular, we define the region of interest (ROI) to be within  spherical Galactocentric radii $r \in [6.5, 9.5]\kpc$ and vertical distances $|z| < 3\kpc$.  In this region, nearly 99\% of all stars belong to the Milky Way's disk, while the remaining $\lesssim 1\%$ belong to the halo and may have been accreted from mergers~\citep{2008ApJ...673..864J}.   Identifying this small fraction of accreted stars is a daunting task given the considerable size of the stellar disk background population.    
A variety of strategies have been explored to achieve this goal, which include looking for structures in action-angle~\citep{Yuksel:2008rf}, energy-momentum~\citep{2000MNRAS.319..657H}, apocenter-pericenter-angular momentum~\citep{Helmi:2005wc}, and/or chemo-dynamic~ \citep[\eg][]{2010A&A...511L..10N, 2016arXiv161100222H, 2018A&A...615A..70P, 2020ApJ...897...39A} space.  

\begin{figure*}[t]
\centering
\includegraphics[width=0.95\textwidth]{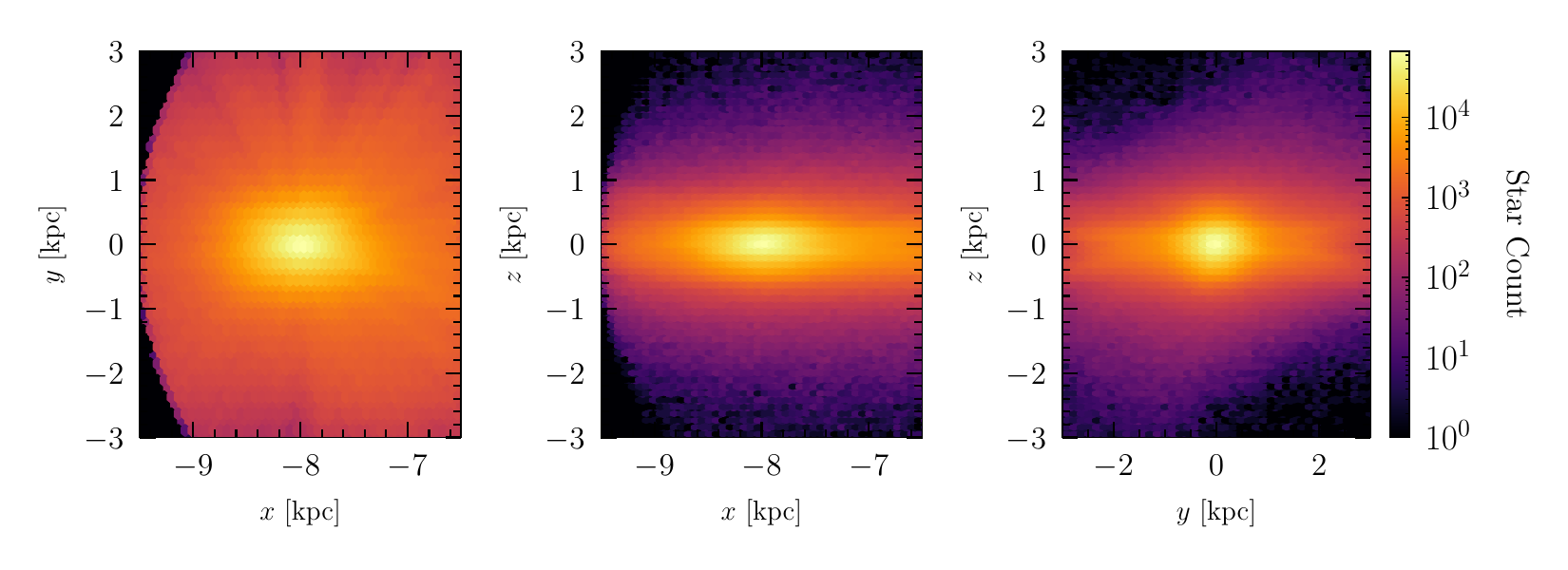}
\caption{Spatial distribution of all \Gaia DR2 stars with parallax $\varpi > 0$, fractional error $\delta \varpi / \varpi < 0.10$, and line-of-sight velocities, falling within the region $r \in[6.5, 9.5]\kpc$ and $|z|< 3\kpc$.  Coordinates are in the Galactocentric Cartesian frame, with $z=0$ the Galactic midplane, $y > 0$ pointing in the direction of the Sun's motion, and $x=0$ at the Galactic Center. The Sun is located at $(x, y, z) = (-8, 0, 0)\kpc$. }\label{fig:selection}
\end{figure*}

The Helmi stream provided the first proof-of-principle that such techniques can be applied successfully~\citep{1999Natur.402...53H}.  First discovered as an overdensity of 13 stars in angular momentum space, the Helmi stream has since been confirmed by a variety of other observations~\citep{2000AJ....119.2843C, 2005A&A...439..551R, 2009ApJ...698..865K, 2009MNRAS.399.1223S, Beers_2017}.  Most recently, it has been studied in \Gaia~DR2~\citep{2018ApJ...860L..11K, 2018A&A...616A..12G} and nearly 600 more potential members were identified using cross matches with spectroscopic surveys~\citep{2019A&A...625A...5K}.  Comparisons with simulations strongly suggest that it is associated with the disruption of a satellite galaxy~\citep{2007AJ....134.1579K}.

The selection criteria typically used to identify accreted stars are conservative and usually reduce potential disk contamination at the expense of excluding a large fraction of stars.  Ideally, one would want to refine the selection process to yield a high-purity sample of accreted stars without sacrificing the overall statistics.  Attempts in this direction have been made using traditional regression and classification techniques~\citep{2019A&A...621A..13V, 2018MNRAS.477.4063M}.  Our focus here, is on the application of deep neural networks to this problem.  In~\citet{ml_paper}, we developed a scheme for using neural networks to successfully distinguish accreted stars from those born in the Milky Way.  We validated this approach through extensive testing on simulated \Gaia mock catalogs~\citep{2018arXiv180610564S} based on the \Latte suite~\citep{Wetzel2016} of the \FIRE simulations~\citep{2015MNRAS.450...53H,2017arXiv170206148H}.  Our method is applied to the subset of \Gaia stars with small parallax errors $\delta \varpi / \varpi < 0.10$, and yields a catalog of likely accreted stars (publicly available in \cite{ostdiek_bryan_2019_3579379}). This work presents the first analysis of a subset of this catalog, focusing on the 4.8 million stars that additionally have radial velocity measurements.

We use several different clustering algorithms to identify velocity structures in this new catalog. We recover known structures such as the Helmi stream and \Gaia Enceladus, clearly showing that the latter extends down to the Galactic midplane.  In addition, we identify several new stream candidates in the ROI.  One of these candidates, which we call Nyx, is a significant contributor in the region studied~\citep{nyx_paper}.  The other two stream candidates, which may be associated with overdensities in~\cite{2018ApJ...860L..11K}, are comprised of $\mathcal{O}(10)$ stars. 

This paper is organized as follows. In \Sec{sec:catalog}, we review the methodology underlying the catalog obtained by \cite{ml_paper}, and explicitly demonstrate how neural network scores correlate with different stellar populations (\eg thin/thick disk and halo).  In \Sec{sec:kinematics}, we perform a Gaussian mixture model analysis to characterize the three most significant structures in the sample: \Gaia Enceladus, Nyx, and the remaining constituents of the stellar halo, which for the remainder of this paper we will refer to as ``Halo.''  We further analyze the data for non-Gaussian structures in \Sec{sec:dbscan}, which is where we recover the Helmi stream, as well as two other stream candidates.  The Appendix includes additional figures that validate the analysis procedure. 

\section{A New Catalog of Accreted Stars}
\label{sec:catalog}
In this section, we take a first look at the catalog of accreted stars.  After briefly summarizing the methodology that underlies the catalog, we provide a number of distributions of the basic properties of the stars identified as accreted and \emph{in situ} by the neural network.  Intriguingly, one can already see hints of novel structures by looking at these distributions.  We will explore these structures more quantitatively in \Sec{sec:kinematics}.

\subsection{Characterizing the Catalog}
\label{sec:sample}

To take maximum advantage of the richness contained within \Gaia DR2, we developed a novel approach utilizing deep neural networks to derive a catalog of accreted stars.  Details and extensive cross-checks of the methodology are presented in~\citet{ml_paper}; we briefly summarize the main points here.  We train the neural network on a combination of simulated mock \Gaia surveys and real measurements of Milky Way stars.  The simulated data is from the \emph{Ananke} mock surveys \citep{2018arXiv180610564S}, based on the \Latte simulation suite~\citep{Wetzel2016,2017arXiv170206148H}.  As we have the full merger history for each simulated galaxy, we can identify stars that are truly accreted, which can be leveraged to both train and validate the networks.  The final network is pre-trained on mock catalogs from three solar positions of the \mi simulated galaxy, using only 5D heliocentric kinematics as inputs (the line-of-sight velocity is not provided to the networks).  We then perform transfer learning on stars within the RAVE~DR5-\Gaia~DR2 cross-matched data~\citep{2017AJ....153...75K} that are identified as accreted with high confidence, \ie, if they are metal-poor $\FeH < -1.5$ and have $|z| > 1.5\kpc$.  Retraining the last layer of the neural network allows it to learn characteristics specific to the Milky Way. 

 The final network is applied to the subset of \Gaia~DR2 stars with parallax $\varpi > 0$ and fractional error $\delta \varpi/ \varpi < 0.10$.  In this work, we further restrict our analysis to only include those stars in the resulting catalog that have measured \Gaia line-of-sight velocities~\citep{2019A&A...622A.205K}, and fall within spherical Galactocentric radii of $r \in [6.5,9.5]\kpc$ and vertical distances $|z| < 3\kpc$ of the Galactic midplane. This results in a final sample size of 4,820,164 stars. Figure~\ref{fig:selection} shows the spatial extent of the sample analyzed here.  Note that \Gaia~DR2 is essentially complete in the optical magnitude range $G \in [7,17]$~\citep{2018A&A...616A...1G}. For $G \in [4,12]$, the radial velocity subset of \Gaia is about 60--80\% complete as compared to the total data set.  Therefore, we do not expect our sample to be spatially complete, and this is evident from close inspection of \Fig{fig:selection}.  For example, the distribution peaks at the Solar circle, and there are rays originating from the Solar position $(x, y, z) = (-8, 0, 0)\kpc$ in the $x - y$ plane. There is also an asymmetry in the $y - z$ plane where the distribution looks almost diagonal, probably due to projection effects (see \cite{2019A&A...622A.205K} for the completeness of the radial velocity subset of \Gaia DR2). Although the stellar sample is not spatially complete, we do expect it to be kinematically unbiased because no selection cuts have been made on velocity.  This is particularly important for the study presented here, as the techniques we use for identifying stellar substructure are based on features in velocity space alone. 

\begin{figure}[t]
\centering
\includegraphics[width=0.47\textwidth]{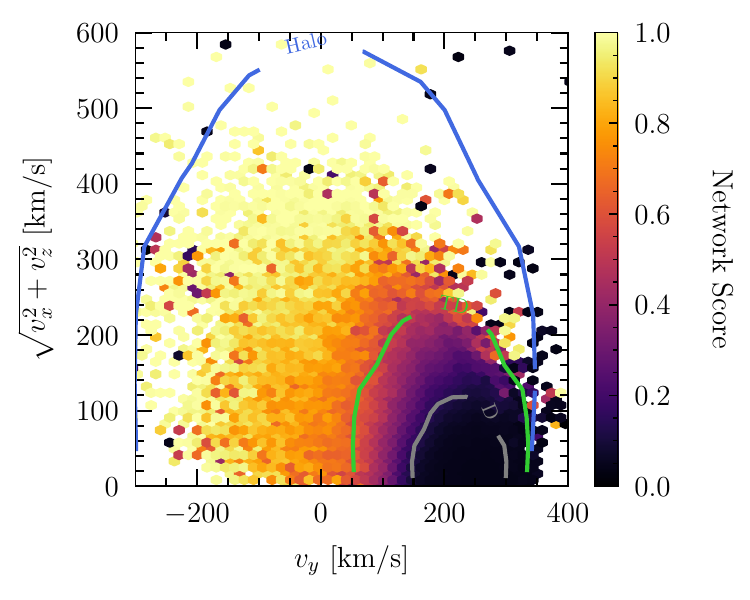}
\caption{Toomre plot of all stars in the subset of \Gaia~DR2 with line-of-sight velocities, $\delta \varpi/ \varpi < 0.10$, and $\varpi > 0$ in the region defined by $r \in [6.5, 9.5]\kpc$ and $|z|<3\kpc$.  Here, positive $v_y$ points in the same direction as the disk rotation.  The colors of the points indicate their network score.  Overlaid are the $3\sigma$ velocity contours of the thin disk~(D), thick disk~(TD), and halo stars, obtained using the model of~\citet{2003A&A...410..527B}.  The network scores are clearly correlated with the different stellar populations. 
}
\label{fig:toomre_scores}
\end{figure}

The neural network gives each star a score that reflects its probability of being accreted, with $S = 0$ indicating a maximum-confidence \emph{in situ} star and $S = 1$ a maximum-confidence accreted star.  Figure~\ref{fig:toomre_scores} shows the Toomre plot in Galactocentric coordinates of the stars in \Gaia DR2 that pass all the selection cuts discussed above.  A gradient of scores is evident, ranging from  stars associated with the thin disk ($S \sim 0$) to stars associated with the stellar halo ($S \sim 1$).  For context, we additionally overlay the best-fit 3$\sigma$ contours for the thin disk, thick disk, and halo stars derived from the model of~\citet{2003A&A...410..527B}. 

Figure~\ref{fig:toomre_scores} clearly shows that the network scores are highly correlated with the expected behavior of these three populations: the lowest scores are associated with the thin disk, while the larger scores are associated with halo stars.  For the most part, thin disk stars have network scores $S \lesssim 0.05$, while thick-disk stars have scores extending up to $S \sim 0.5$\footnote{Running the neural network on the \texttt{Galaxia} \citep{2011ApJ...730....3S} simulation demonstrated that the thick-disk component is the hardest one to score as accreted or \emph{in situ}. In actuality, the Milky Way's thick disk might be a combination of the two; see~\cite{2013A&ARv..21...61R} and references therein.} --- see \Fig{fig:disk_hists} for Toomre plots of stars with scores $S < 0.05$ and $S \in [0.3, 0.5]$.  For values of $v_y > 200\kms$ and $\sqrt{v_x^2 + v_z^2} \sim 100$--$200\kms$, stars that appear to belong to the thick disk according to the~\citet{2003A&A...410..527B} model have very low network scores, close to 0. These stars are easily identified by the network as being \emph{in situ} because their $v_y$ velocities are large and positive. Note that due to the focus of the training regimen, the network score reflects the probability that a star is accreted, and as such it does not have any information on the different components of the disk.\footnote{It is in principle possible to use multi-class classification to identify different stellar populations, but that is outside the scope of this work.}  It is therefore rather remarkable that the network scores do correlate with known stellar populations.

\begin{figure*}[t]
\begin{center}
\includegraphics[width=0.90\textwidth]{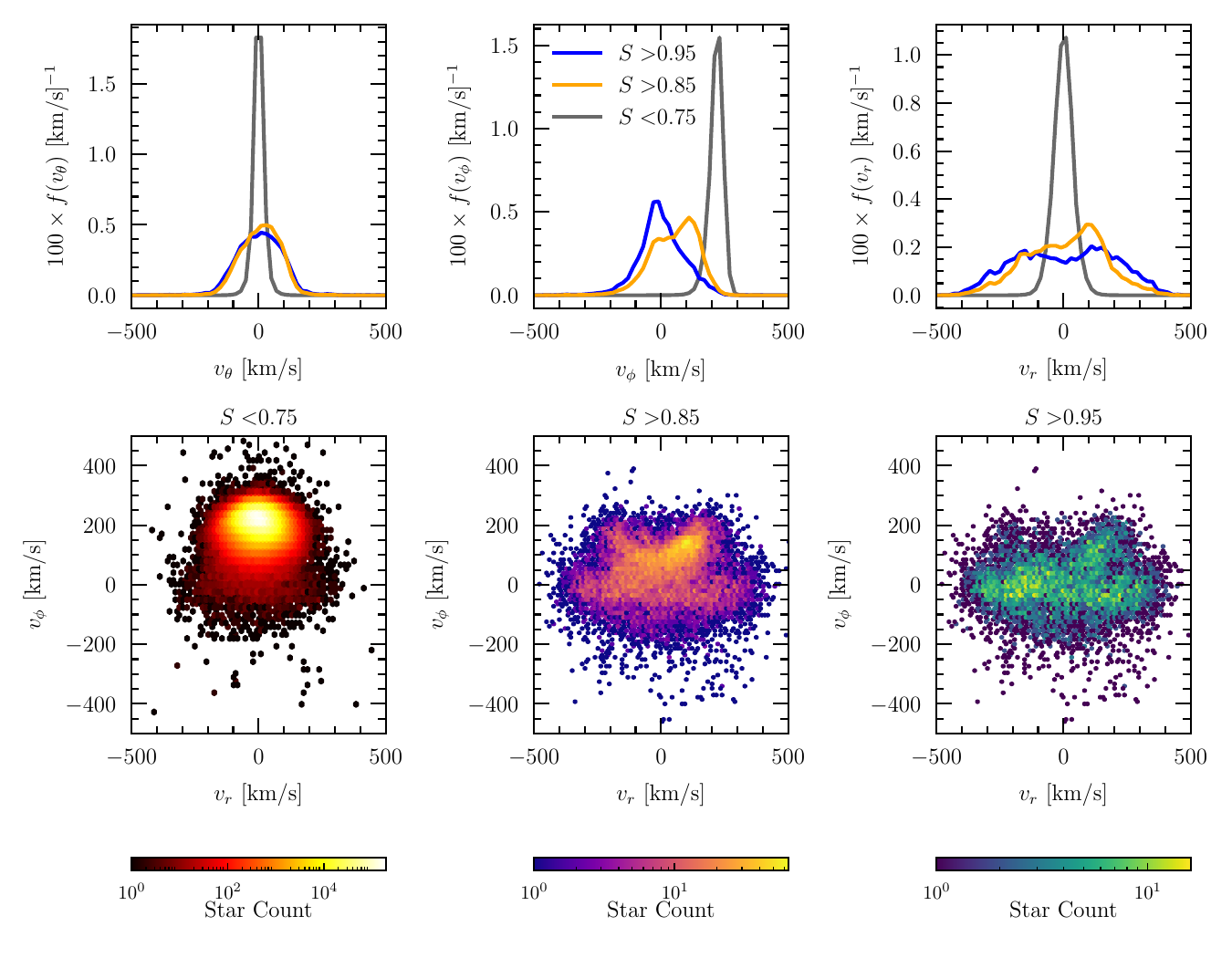}
\caption{The top row shows the spherical Galactocentric velocity distributions of the stars in the \Gaia DR2 dataset that pass the cuts described in \Sec{sec:sample}.  Here, $v_\phi$ rotates with the disk of the Milky Way and positive $v_r$ points towards the Galactic Center.  Distributions are provided for the \emph{in situ} ($S<0.75$), canonical  ($S > 0.85$), and high-purity ($S > 0.95$) samples.
The second row provides $v_r - v_\phi$ correlation plots for these three samples.  The \emph{in situ} sample is clearly peaked at the disk rotation ($v_\phi \sim 220\kms$).  \Gaia Enceladus is apparent in the canonical and high-purity samples as the radial overdensity at $v_\phi \sim 0\kms$.  As we will demonstrate, the stellar overdensity at  $v_r \sim 130\kms$ and $v_\phi \sim 130\kms$, which is most apparent in the high-purity sample (but also present in the canonical sample), corresponds to a new stream, which we call Nyx.   }
\label{fig:KinematicRV_AllStars}
\end{center}
\end{figure*}

To create a concrete dataset of accreted stars to work with, we must choose a cut on the network scores.  In~\cite{ml_paper},  we developed a principled way of identifying this cut by analyzing the network performance on mock catalogs, where truth information is available.  Our criteria was to minimize the difference between velocity distributions of the stars that are selected and the distributions of all of the \emph{truth-level} accreted stars.   In particular, we found that a cut of $S\gtrsim 0.75$ least biased the distributions of the accreted stars.\footnote{All stars in \Gaia with a score greater than 0.75 are made available upon request from the corresponding authors.}  Here, we will use more restrictive cuts on the score to improve the purity of the sample and facilitate the Gaussian Mixture Model analysis described in \Sec{sec:clusters}.  Our \emph{canonical} sample will be defined with $S>0.85$; it contains 22,296 stars total.  
We additionally explore the impact of placing a much more restrictive cut of $S > 0.95$ to create a high-purity sample (9,379 stars total).  The high-purity sample minimizes disk contamination at the expense of biasing the velocity distributions, while the canonical sample reduces the purity by introducing some disk stars, but should produce less biased distributions.

\begin{figure*}[t]
\begin{center}
\includegraphics[width=0.45\textwidth]{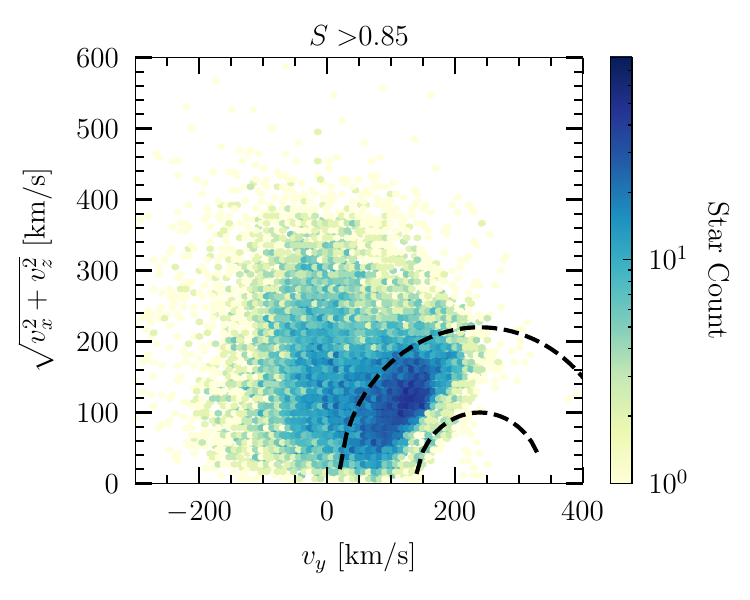}
\qquad
\includegraphics[width=0.45\textwidth]{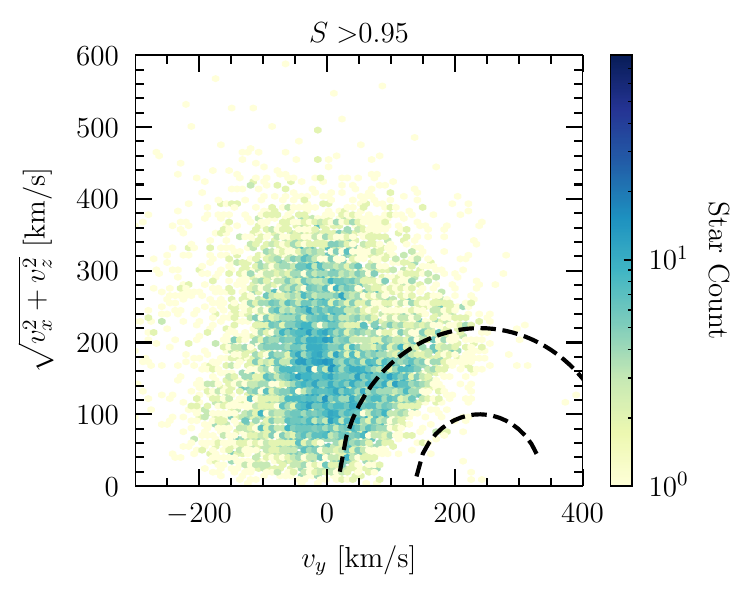}
\caption{Toomre plots for the stars in the accreted catalog with the canonical (left) and high-purity (right) sample.
The black, dashed lines show velocities within 100 and 220 km/s of the local frame.  The network does not classify any stars that are within the inner dashed circle as accreted, and it has higher confidence that stars that are further away from the local standard of rest are accreted. 
}
\label{fig:ToomreKinetic}
\end{center}
\end{figure*}

\subsection{Accreted Versus In situ Stars}
We now explore the kinematic distributions of the canonical and high-purity samples.  The top row of \Fig{fig:KinematicRV_AllStars} shows the velocity distributions in spherical Galactocentric coordinates ($v_r$, $v_\phi$, $v_\theta$) for three different cuts on the network score: $S<0.75$ (\emph{in situ}), $S>0.85$ (canonical accreted), and $S > 0.95$ (high-purity accreted).  The narrow peak near $v_\phi \sim 200\kms$ for the stars with $S< 0.75$ shows that the network clearly classifies stars whose rotations are consistent with the disk as \emph{in situ}.  The $v_r$ and $v_\phi$ distributions differ between the canonical and high-purity sample.  While both are  roughly bimodal in $v_r$, the canonical sample has a smaller dispersion. The canonical sample has a stronger peak at $v_\phi \sim 150\kms$, which may due to some disk contamination.

To further explore the differences between the canonical and the high-purity cuts, we provide the  $v_r- v_\phi$ distributions of the resultant datasets in the bottom row of \Fig{fig:KinematicRV_AllStars}. The \emph{in situ} sample ($S<0.75$) is dominated by the stars at $v_r \sim 0 \kms$ and $v_\phi \sim 200\kms$ as expected, although it also includes some accreted stars that happen to be scored below $S=0.75$ by the network. Both the high-purity and canonical samples show an elongated structure in the radial direction that spans $v_r \in [-400, 400] \kms$, with $v_\phi \sim 0\kms$.  This is \Gaia Enceladus, further discussed in \Sec{sec:structures}.  When we implement the canonical cut, we find that the network carves out a nearly circular region in this parameter space, because any star that falls within this region is very likely to be \emph{in situ}.  It does however leave a distinct half-moon structure in this plane, particularly apparent in the middle panel of \Fig{fig:KinematicRV_AllStars}, second row. If we increase the cut to $S > 0.95$, so that we only plot the stars that the network labels as accreted with high confidence, we find that the half-moon structure dissolves into a localized overdensity at $v_r \sim 150\kms$ and $v_\phi \sim 140 \kms$.  We call this new structure Nyx, and will describe it in more detail in \Sec{sec:structures}. It is important to note that Nyx is present in the canonical sample as well, but is much easier to identify in the high-purity sample.  

Figure~\ref{fig:ToomreKinetic} provides the Toomre plots for the canonical and high-purity cuts.  
The black-dashed lines are centered around $v_\text{lsr} = 220\kms$ and extend to $|v - v_\text{lsr} | > 100 - 220\kms$. 
As with the $v_r - v_\phi$ distributions shown in \Fig{fig:KinematicRV_AllStars}, the network cuts out the region within the inner dashed circle.  A common selection criterion for identifying accreted stars is to require that $|v - v_\text{lsr} | > 180$--$220\kms$ (see~\cite{2018A&A...615A..70P} for a review).  By default, such a cut ignores stars that fall within (approximately) the outermost dashed circle in \Fig{fig:ToomreKinetic}, biasing the resulting distribution of stars in velocity space.  This consequently ignores any substructure that co-rotates with the disk.  Our deep-learning--based catalog has been constructed to minimize such a bias, thereby enabling us to look for prograde structures.  Several structures stand out in both the canonical and high-purity samples.  The vertically extended overdensity centered at $v_y \sim 0\kms$ is \Gaia Enceladus.  Second, the overdensity centered at  $v_y \sim 130\kms$ and $\sqrt{v_x^2 + v_z^2} \sim 130 \kms$ is Nyx.   Nyx becomes more evident in the high-purity sample where the disk contamination is reduced.

We emphasize that the kinematic distribution of the stars that are easiest to label as accreted are not guaranteed to be representative of the distribution for \emph{all} accreted stars.  Therefore, moving forward, we will present results for both the canonical and high-purity samples, although we emphasize the possibility of biases in the distributions, especially as the cut on $S$ becomes more restrictive~\citep{ml_paper}.

\section{Dominant Kinematic Structures}
\label{sec:kinematics}

In this section, we focus on characterizing the largest (in terms of overall star count) kinematic structures present in the sample of accreted stars.  We will use a Gaussian mixture model approach to extract three distinct features, including the new stream Nyx.

\subsection{A Structure Finding Algorithm}
\label{sec:clusters}

 To obtain a quantitative estimate for the number of components in the high-purity accreted dataset (defined by the sample of \Sec{sec:sample} with $S>0.95$), we run a Gaussian mixture model using \texttt{scikit-learn}~\citep{scikit-learn} on the selected stars in Galactocentric velocity space.  This first pass ignores the impact of finite measurement errors on the stellar velocities.  We allow for up to 9 Gaussian distributions, and evaluate how many are necessary using the Bayesian Information Criterion (BIC). The BIC decreases rapidly as the number of Gaussians is increased to 4, after which it stabilizes.  Note that \Gaia Enceladus is best characterized by two Gaussians with nearly identical means and dispersions in $v_\phi$ and $v_\theta$ that are centered at equal but opposite values of $v_r$ \citep{2019ApJ...874....3N}.  Hence, the fact that 4 Gaussians are required to minimize the BIC tells us that the dataset contains three dominant kinematic structures.  We will refer to these three structures as: Enceladus, Nyx, and the Halo.  We emphasize that these three components taken together comprise the totality of the accreted stellar halo in our model.\footnote{There is possible disk contamination even in the high-purity sample. However these stars are usually not representative of the \emph{in situ} sample, and therefore we do not expect them to form a distribution that can be picked up by the Gaussian mixture model.}  In other words, the component labeled as ``Halo'' is comprised of accreted stars in the stellar halo (some of which might be in coherent structures) that are not resolved by the mixture modeling.

\begin{figure*}[t]
\centering
\includegraphics[width=0.95\textwidth]{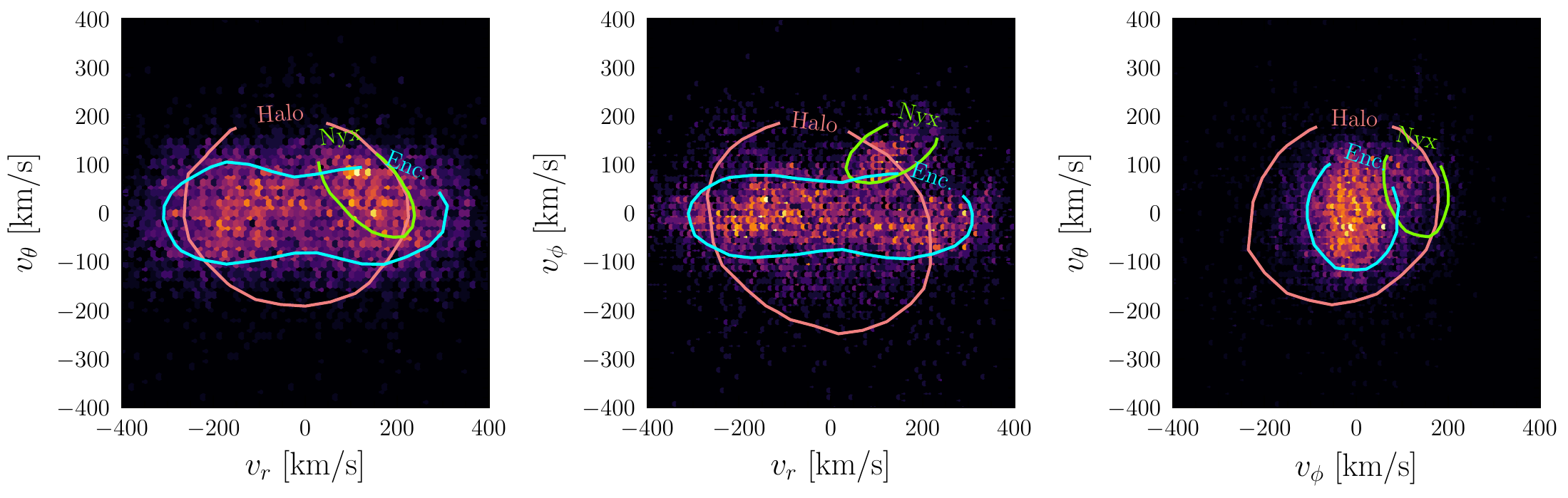}
\includegraphics[width=0.95\textwidth]{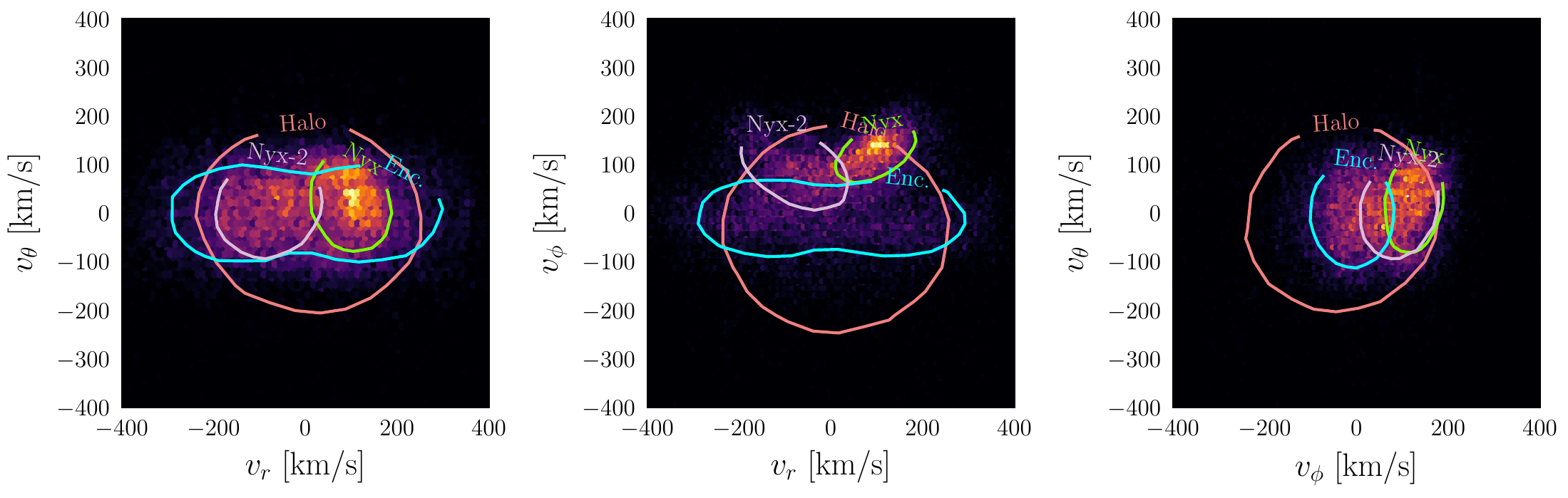}
\caption{Kinematic distributions of stars in the \Gaia~DR2 catalog that have measured radial velocities and fall within Galactocentric radii $r \in [6.5, 9.5]\kpc$ and vertical distances $|z| < 3\kpc$ of the midplane.  These stars have been identified as accreted by the neural network developed in~\cite{ml_paper}; the top row shows the distributions for the high-purity sample ($S>0.95$) while the bottom row shows the distributions for the canonical sample  ($S>0.85$). The two-dimensional distributions are shown for the Galactocentric velocity coordinates $v_r$, $v_\theta$, and $v_\phi$.  In the high-purity sample, a Gaussian mixture study recovers a Halo component with large dispersion (pink), \Gaia Enceladus (blue), and Nyx (green).  These same components are also identified in the canonical sample, which additionally includes  
a separate prograde stream, that we refer to as Nyx-2 (purple).  
 Note that ``Halo'' refers to the remaining accreted stars in the stellar halo that are not individually resolved by the mixture analysis.  }
\label{fig:all_structure}
\end{figure*}

\begin{figure*}[t]
\centering
\includegraphics[width=0.95\textwidth]{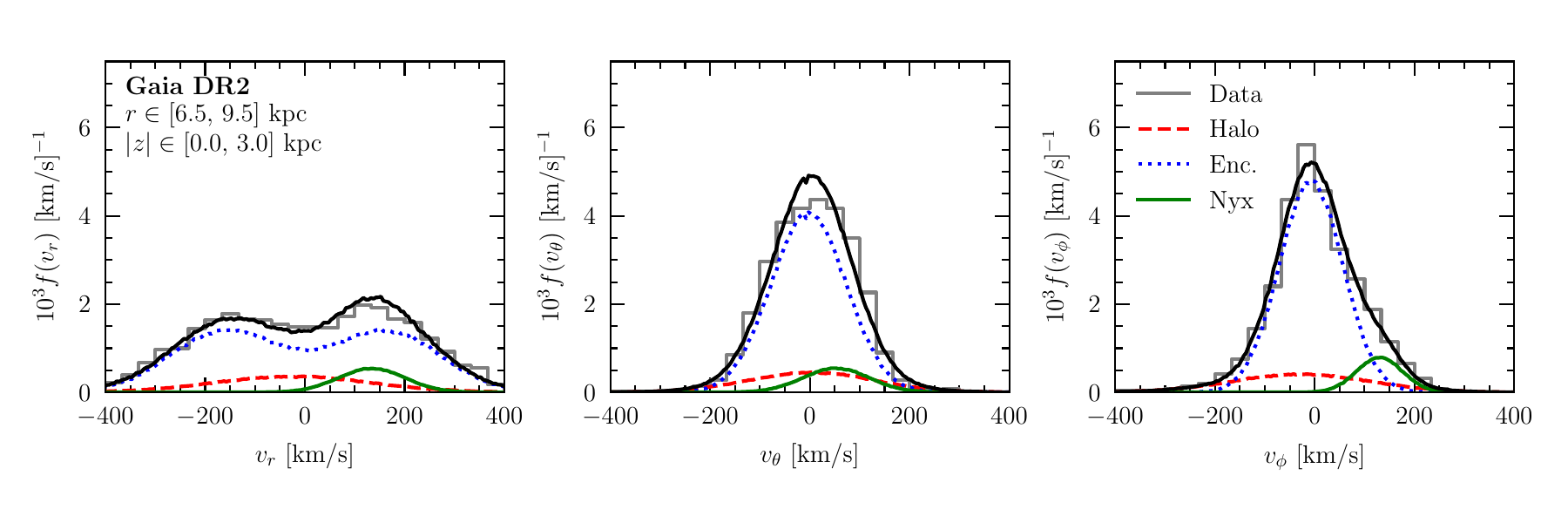}
\includegraphics[width=0.95\textwidth]{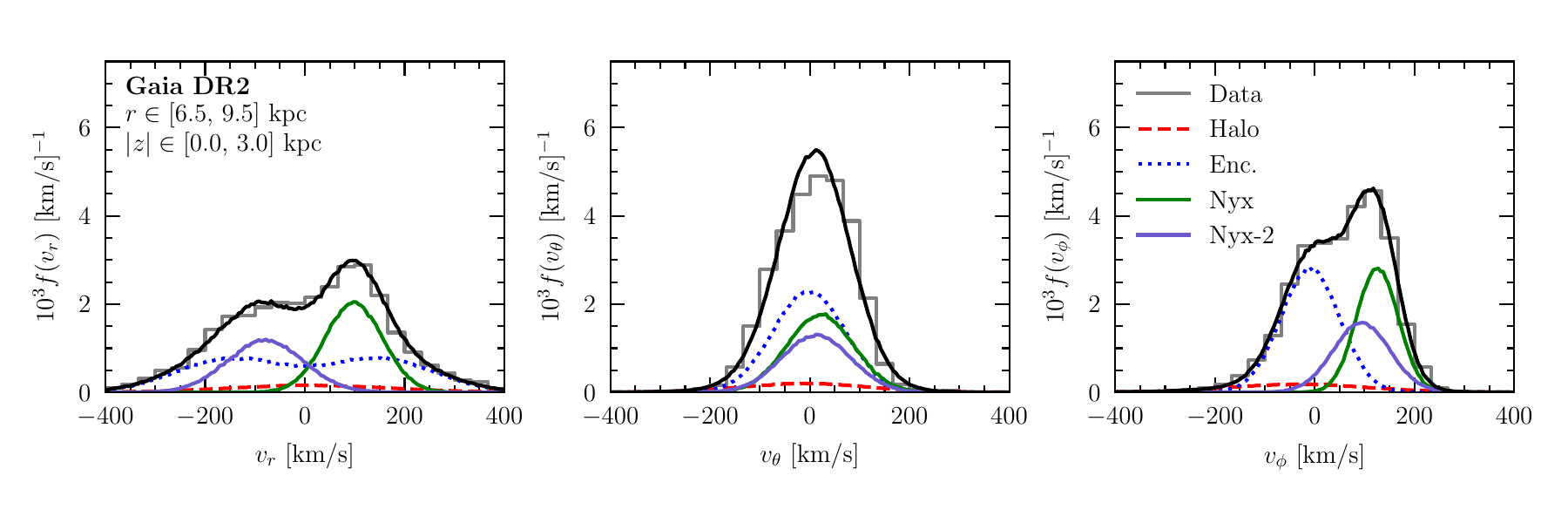}
\caption{Best-fit velocity distributions for all stars with network scores $S > 0.95$ (top row) and  $S>0.85$ (bottom row), which have measured radial velocities and fall within Galactocentric radii $r \in [6.5, 9.5]\kpc$ and vertical distances $|z| < 3\kpc$ of the midplane.  The distributions are shown for the Halo (red dashed), Enceladus (blue dotted), and Nyx (green solid) components.  The total distribution of the model is shown in solid black, and can be compared to the data (gray histogram). In the canonical sample  ($S>0.85$), we also identify a second prograde stream, called Nyx-2. A plot of the model residuals for the high-purity sample is provided in \Fig{fig:residuals}. }
\label{fig:gmm_modeling}
\end{figure*}

The simple mixture analysis that yielded the BICs on the high-purity sample is efficient, but does not account for the uncertainties in the measured parameters of the stars. The preliminary analysis informs the number of Gaussian distributions we include moving forward. To properly account for the errors, we perform our own dedicated study to derive the best-fit parameters of the velocity distributions of these separate components.   We define the likelihood that a star $i$ belongs to either the Halo or Nyx by 
\begin{equation}
p_k \big(O_i \, \big|\, \theta \big) = \CN \left( \vec{v}_i \,\middle| \, \boldsymbol{\mu}^k, \boldsymbol{\Sigma}^k_i \right),
\label{eq:likelhood}
\end{equation}
where $k = h$ (Halo) or $n$ (Nyx), $O_i = \{v_{r,i}, v_{\theta,i}, v_{\phi,i} \}$ are the velocities of star $i$ in spherical Galactocentric coordinates, and $\theta$ is the set of free model parameters.  $\CN$ denotes the multivariate normal distribution with mean $\boldsymbol{\mu} = (\mu_r, \mu_\theta, \mu_\phi)$ and covariance  $\boldsymbol{\Sigma}_\text{true}$, which is a function of the dispersions $\sigma_{r, \theta, \phi}$ and the correlation coefficients $\rho_{r\theta, r \phi, \theta \phi}$. The measurement errors are taken into account as $\boldsymbol{\Sigma}_i^k = \boldsymbol{\Sigma}_\text{true}^k + \boldsymbol{\Sigma}_{\text{err},i}$, where $\boldsymbol{\Sigma}_{\text{err},i}$ varies from star to star but does not depend on the model parameters.
The likelihood for Enceladus is the sum of two Gaussians with opposite means in the radial direction:
\begin{equation}
p_e \big(O_i \, \big|\, \theta \big) =  \frac{1}{2} \, \Big[ \, \CN \Big( \mathbf{v}_i \Big|\, \boldsymbol{\mu}^{e}, \boldsymbol{\Sigma}^e_i \Big) +  \CN \Big( \mathbf{v}_i \Big|\, \boldsymbol{\mu}^{\tilde{e}},  \boldsymbol{\Sigma}^e_i \Big)  \,\Big] \,  ,
\label{eq:sublikelihood}
\end{equation}
where $\boldsymbol{\mu}^{\tilde{e}} = (-\mu_r, \mu_\theta, \mu_\phi)^{e}$.  The total likelihood is therefore defined as 
\begin{equation}
p\big(\{ \mathcal{O}_i \} \big| \theta\big) = \prod_{i=1}^N \sum_{j=h, n, e} Q_j   \, p_j\, \big(O_i \, \big|\, \theta \big)    \, ,
\label{eq:TotalLikelihood}
\end{equation}
where $Q_j$ is the fractional contribution of each component (constrained to add up to 1 over all $j$).  This type of likelihood analysis is very similar to what was performed in~\citet{2019ApJ...874....3N}, except that here we only cluster in kinematic space, and do not include metallicity information in the likelihood.  

\begin{table}[b]
\footnotesize
\begin{center}
\begin{tabular}{C{1.2cm}C{1.2cm}C{1.5cm}C{1.4cm}C{1.5cm}}
\Xhline{3\arrayrulewidth}
\renewcommand{\arraystretch}{1}
&\multicolumn{4}{c}{Priors} \\
& Halo & Enceladus & Nyx & Nyx-2 \\ 
\hline
$\mu_r$&  $[-70,70]$ & $[0, 250]$  & $[-70, 200]$ & $[-300,200]$ \\
$\mu_\theta$&  $[-70,70]$ & $[-200,200]$ & $[-70,70]$ & $[-70,70]$\\
$\mu_{\phi}$&  $[-70,70]$ & $[-200, 200]$& $[0,300]$ & $[0,300]$ \\
$\sigma_{r, \theta, \phi} $& $ [0,400]$ & $[0, 400]$& $[0, 400]$ & $[0,400]$\\
$\rho_{{r \theta}, {r \phi}, {\theta \phi}}$&  $[-1,1] $ & $[-1,1]$ & $[-1,1]$  & $[-1,1]$\\
$Q$&  --- & $[0, 1]$ & $[0,1]$ &  $[0,1]$ \\
\Xhline{3\arrayrulewidth}
\end{tabular}
\caption{Parameters and associated prior ranges for the halo, \Gaia Enceladus, Nyx, and Nyx-2.  All priors are taken to be linear.  Note that the Nyx-2 component is only included when analyzing the canonical  ($S>0.85$) sample.}
\label{tab:priors}
\end{center}
\end{table}

We also present results using the  $S>0.85$ sample. In this case, the total likelihood in \Eq{eq:TotalLikelihood} includes  an additional Gaussian, modeled following \Eq{eq:likelhood},  intended to capture a second component of Nyx at negative radial velocity, which we  refer to as Nyx-2.\footnote{Using the same technique of evaluating the BICs, we find that the number of preferred Gaussians stabilizes after  6.  We choose to include 5 Gaussians (two of which correspond to \Gaia Enceladus). Adding a sixth Gaussian leads to two overlapping distributions in velocity space, roughly at the location of Nyx.}  

We run \texttt{emcee}~\citep{2013PASP..125..306F} to find the posterior distributions for the free parameters of the separate components.  With nine free parameters for each population, and two additional parameters to quantify the relative fractions, this is a 29 parameter fit for the $S > 0.95$ sample study. With  4 components, and  3 relative fractions, the  $S > 0.85$ sample requires  39 free parameters. The priors are linear, and their ranges are listed in \Tab{tab:priors}. We use  200 walkers,  8000 steps for the burn-in stage, and   1000 steps for the analysis of the high-purity (canonical) samples. The relevant corner plots are provided in the Appendix as Figs.~\ref{fig:ratios_corner}--\ref{fig:subs_corner_75}.

\vspace{0.5in}
\subsection{Properties of Enceladus and Nyx}
\label{sec:structures}

\begin{figure*}[t]
\centering
\includegraphics[width=0.45\textwidth]{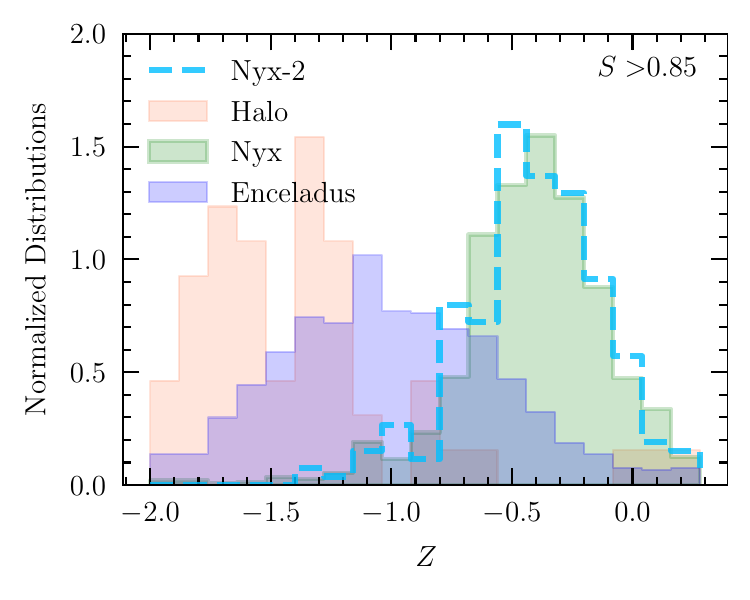}
\qquad
\includegraphics[width=0.45\textwidth]{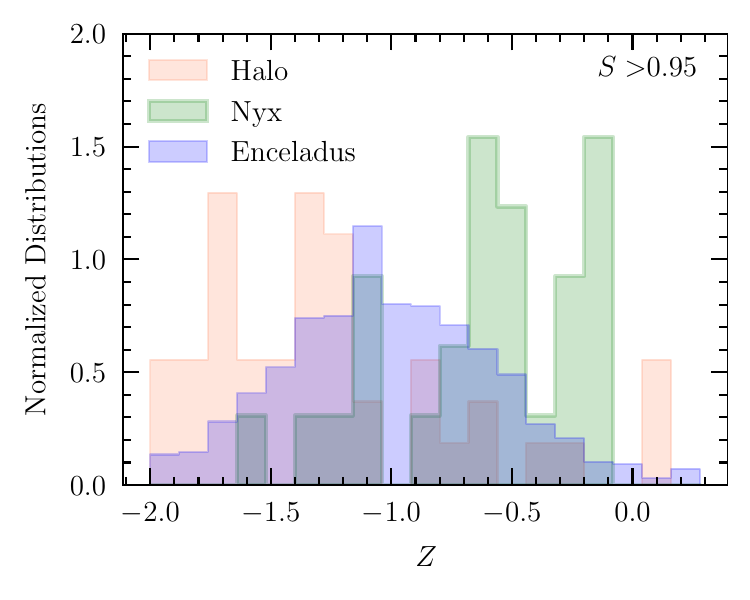}
\caption{ {updated} The metallicity $Z$ distributions for the Halo, Nyx, and Enceladus components in the canonical (left) and high-purity (right) samples. The use of $Z$ here should not be confused with $z$, the vertical distance above the plane of the Milky Way. We also show the distribution for Nyx-2, which is only identified in the canonical sample.  A star is associated with any one of these populations if it has a $>95\%$ ($>85\%$) probability of belonging to the respective Gaussian component for the high-purity (canonical) sample. All distributions are normalized to unity.  Note that the relative normalizations between components reflect the fractions within our sample. Metallicities are taken from~\cite{2018MNRAS.481.4093S}. 
}
\label{fig:Z_distributions}
\end{figure*}

In this section, we investigate the properties of the dominant kinematic structures in the catalog. The top (bottom) row of \Fig{fig:all_structure} provides the 2D kinematic distributions of the stars in the $S > 0.95$ ( 0.85) catalog in Galactocentric spherical coordinates. The lines show the 68\% contours of the posterior distributions for the separate stellar components, modeled using the likelihood in \Eq{eq:TotalLikelihood}.  The corresponding 1D distributions for the posteriors are shown in \Fig{fig:gmm_modeling}.  

\Gaia Enceladus is the radial bi-modal population in \Fig{fig:all_structure} (blue line).  By construction, the lobes of Enceladus have the same mean and dispersion in $v_\theta$ and $v_\phi$ and are located at equal, but opposite, values of radial velocity, see \Eq{eq:sublikelihood}.  The general properties of the Enceladus distribution are largely consistent with results from a previous study using the SDSS-\Gaia cross match~\citep{2019ApJ...874....3N}.
From the radial distributions shown in \Fig{fig:gmm_modeling}, we see that the peaks are located closer together in the canonical sample than in the high-purity sample. These differences may be a result of kinematic biases that are introduced as the score cut is increased.  Additionally, as shown in \Fig{fig:sausage_spatial}, we find that stars with a high probability of being associated with Enceladus clearly extend down to the Galactic midplane. This corroborates the hypothesis that the Enceladus merger contributes to the local dark matter distribution~\citep{Necib:2018igl,2019ApJ...874....3N}.

Nyx is the prograde group of stars characterized by its significant radial velocity (\Fig{fig:all_structure}, green line).  In particular, Nyx moves in the same direction as the Galactic disk, but its rotational velocity lags by  $\sim 90\kms$.  Its radial velocity distribution has a mean value of  $134\kms$ and dispersion of  $\sim 67\kms$.  In the canonical sample, we find a corresponding structure, which we call Nyx-2, that has an average   $v_\phi$ of $94\kms$, compared to the Nyx value (in the same sample) of  $125\kms$, but equal and opposite $v_r$ of  $-79\kms$ compared to  $98\kms$ for Nyx. 

 The fraction of Nyx stars is 9\% and 30\% in the high-purity and canonical samples, respectively.  Nyx-2 corresponds to  22\% of the canonical sample.  Enceladus is the dominant structure in both samples, comprising  77\% (42\%) of the high-purity (canonical) set.  The relative fraction of Enceladus is reduced in the canonical sample primarily because of the presence of Nyx-2.

A complete discussion of the properties of Nyx and Nyx-2, and their potential origin, is presented in~\cite{nyx_paper}.  Briefly summarizing what is detailed there, Nyx is a coherent stellar stream on an eccentric orbit ($e \sim 0.6$) whose distribution is highly unlikely to be drawn from the expected smooth distribution of the thick disk.  Its behavior is consistent with prograde streams observed in simulations, which are created when a massive satellite is dragged into the disk plane by increased tidal friction~\citep{1986ApJ...309..472Q, 1989AJ.....98.1554L, 1996ApJ...460..121W,  2003ApJ...597...21A, Read:2008fh, 2009MNRAS.397...44R, 2009ApJ...703.2275P, 2010JCAP...02..012L, Pillepich:2014784}.  The kinematics of Nyx-2 suggest that it is related to Nyx, and may be tidal debris from a separate passage of the same satellite.  We approach this conclusion  cautiously only because the network scores for Nyx-2 stars are generally lower than those for Nyx and its detection is thus not as robust.    

To study the chemical abundances of each component, we use the cross match of stars provided by~\citet{2018MNRAS.481.4093S}, covering the spectroscopic catalogs APOGEE~\citep{2017AJ....154...94M}, LAMOST~\citep{2012RAA....12..735D}, RAVE~\citep{2006AJ....132.1645S}, GALAH~\citep{2015MNRAS.449.2604D}, Gaia-ESO~\citep{2012Msngr.147...25G}, and SEGUE~\citep{2009AJ....137.4377Y}.  
Out of the  9,379 (22,296) stars in the high-purity (canonical) sample, we find cross matches for  4,255 (10,309). 
We histogram each star's metallicity $Z$ in \Fig{fig:Z_distributions}.  A star is associated with a given component if the mixture analysis finds that it has a probability greater than 95\% of belonging to it in the high-purity or canonical samples. 
The Halo  has a mean at $Z \sim -1.3$ in the high-purity sample, with large dispersion and a long tail towards more metal-poor stars.  Fewer stars are associated with the Halo in the canonical sample as compared to the high-purity sample (less than  3\% versus  10\%), but the metallicities are largely consistent between the two. Enceladus also has a large dispersion in $Z$, and peaks at  $Z\sim -1$ in both the high-purity sample and the canonical samples. 
 Nyx stars however have a much narrower $Z$ distribution, peaking at  $Z \sim -0.6$ ($Z \sim -0.4$) in the high-purity (canonical) sample.  In both the canonical and high-purity samples, the distributions of the different components are consistent. Interestingly, the distribution of Nyx-2 in the canonical sample matches that of Nyx, strengthening the argument that they are debris from the same progenitor. Spectroscopic follow ups will be crucial to more deeply understand the origin of Nyx and Nyx-2.

\begin{figure}[t]
\centering
\includegraphics[width=0.45\textwidth]{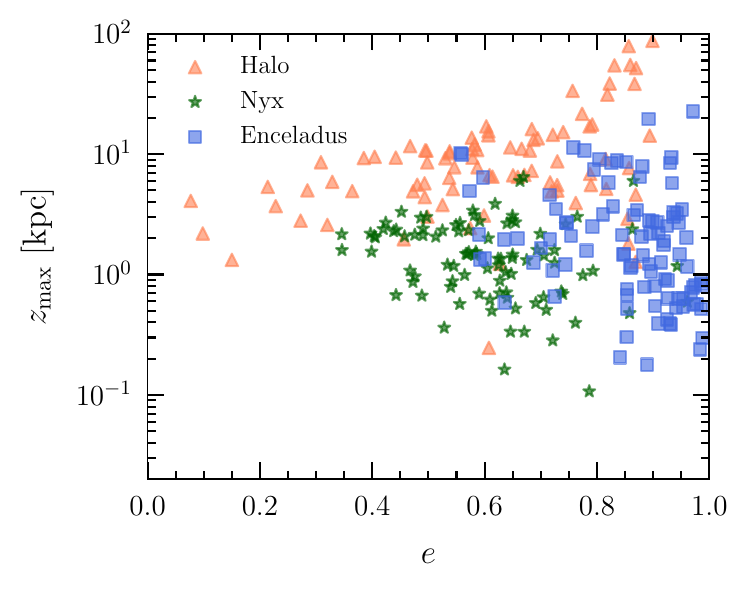}
\caption{  Scatter plot of the maximum vertical distance versus eccentricity for Nyx, Halo, and \Gaia Enceladus stars in the high-purity sample.  We only show a random selection of 100 stars for the Halo and \Gaia Enceladus. Nyx stars are typically concentrated in the disk plane.  However, their eccentricities are larger than expected for thick-disk stars~\citep{2017ApJ...850...25L}. }
\label{fig:e_zmax_all}
\end{figure}

In \Fig{fig:e_zmax_all}, we provide a scatter plot of the maximum vertical distance from the midplane, $z_\text{max}$,  versus the eccentricity, $e$, of the different components as obtained from a simple calculation of their orbits. These parameters are evaluated when running the stellar orbits back in time 1  Gyr over 1000 steps, using the package \texttt{gala} \citep{gala}. To take the measurement errors into account, we resample each star 100 times from the position and velocity Gaussians, and average their eccentricities and $z_\text{max}$. Figure~\ref{fig:e_zmax_all} only shows 100 random stars from each component to not overwhelm the plot. Full distributions of the eccentricities, $z_\text{max}$, apocenters, and pericenters of each component are provided in the Appendix, see Figs.~\ref{fig:orbit_nyx}--\ref{fig:orbit_sausage}.

\begin{figure*}[t]
\centering
\includegraphics[width=0.95\textwidth]{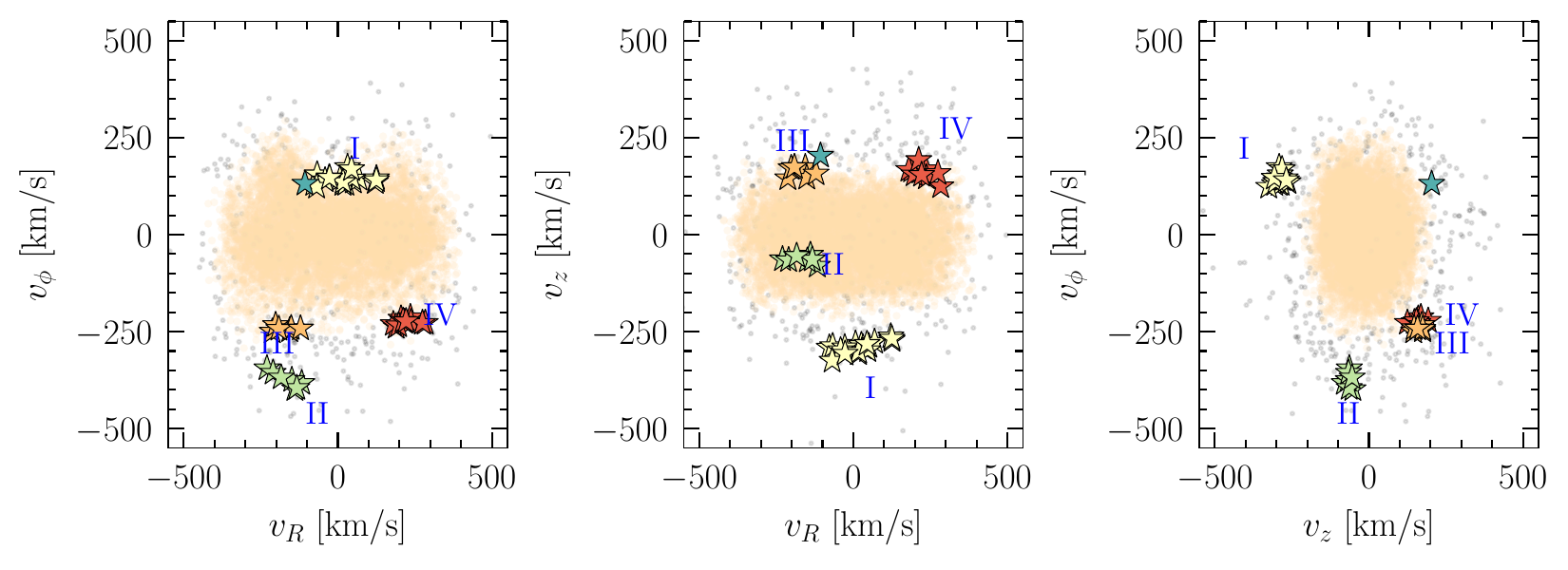}
\caption{Results of running the DBSCAN clustering algorithm on the high-purity ($S > 0.95$) accreted stars in the catalog for a single realization.  The clusters are shown in Galactocentric cylindrical coordinates ($v_R$, $v_\phi$, $v_z$).  The colored stars indicate groups of stars that are associated kinematically; the gray dots are outliers identified by the method. The large pale orange cluster of points corresponds to the Enceladus and Nyx structures, which DBSCAN associates together.   We repeat this procedure 100 times, sampling over the errors in stellar position and velocity.  Groups I-IV are the most robust as they are identified in over  70 of the 100 iterations. The green star without an associated number  is an example of a cluster that is not present through more than 70 iterations. }
\label{fig:dbscan}
\end{figure*}

Enceladus stars have largely eccentric orbits ($e > 0.8$), confirming that the satellite progenitor was on a highly radial orbit~\citep{2018ApJ...862L...1D}.  Similarly, Halo stars have large eccentricities and extend up to $\sim 100\kpc$ above the plane, on average higher than Enceladus.  The 1D histograms of the eccentrities show that Enceladus stars are peaked towards larger eccentricities (\Figs{fig:orbit_halo}{fig:orbit_sausage}).   Nyx stars extend to   $z_\text{max} \sim 0.1$--$4\kpc$ above the plane at eccentricities  $e \sim 0.3$--$0.8$ --- the latter are higher than what is expected of thick-disk stars \citep{2017ApJ...850...25L}. 

\section{More Subtle Kinematic Structures}
\label{sec:dbscan}

In \Sec{sec:kinematics}, we discussed the three most dominant structures found in the high-purity and canonical catalogs. To see if there is any evidence for non-Gaussian structures, we now use the algorithm \texttt{DBSCAN}\footnote{\url{https://scikit-learn.org/stable/modules/clustering.html}} \citep{scikit-learn}. This clustering algorithm is designed to identify connected overdensities; we will apply it to the 3D velocities in Galactocentric cylindrical coordinates, not spherical as in \Sec{sec:kinematics}, which will allow for easier comparison with previous studies. Note that \texttt{DBSCAN} does not require any assumptions regarding the shape of the underlying distributions, and therefore can identify non-Gaussian structures. This is particularly important when hunting for small stream-like features. \texttt{DBSCAN} depends on the minimum number of stars in a group, $n_\text{min}$, and the compactness parameter, $\epsilon$. 

It is important to take into account the errors on the positions and velocities of the stars while using \texttt{DBSCAN}. To do so, we resample the stars over their errors and run \texttt{DBSCAN} 100 times, identifying the groups for each independent run.  In \Fig{fig:dbscan}, we show the resulting groups for $n_\text{min} = 5$ and $\epsilon = 0.4$ for one such realization, which returned nine groups.  Note that one of these groups (shown in pale orange) is attributed to the combination of \Gaia Enceladus and Nyx, as their distributions overlap and they are thus not separated by this algorithm. 

Each realization of \texttt{DBSCAN} is prone to statistical fluctuations as we vary the stars within their measurement errors. To find the most robust structures, we save the centers of each of the groups  identified in each the 100 resamplings, and then run  \texttt{DBSCAN} over these 100 realizations to find clusters of group centers that are present in more than  70 realizations, with $\epsilon = 0.3$ and $n_\text{min} = 50$. We find  four stream candidates that pass these cuts for the high-purity sample, see \Tab{tab:small_structures}. Interestingly, when running the same algorithm on the canonical sample, the same three groups pass these requirements. 

The stream that is the most robust --- in that it is found by all 100 realizations --- is an overdensity at  $(v_R, v_{\phi}, v_z) = (21, 141, -286)$ km/s, shown in the realization of \Fig{fig:dbscan} as the stars in  yellow (labeled as Group~I). This overdensity can be matched to the Helmi stream~\citep{1999Natur.402...53H},\footnote{It also overlaps the S2 stream in \cite{2018MNRAS.475.1537M}, which is believed to be related to the Helmi stream.} which has been reported in \Gaia~DR2 by several studies, including~\cite{2018A&A...616A..12G, 2018ApJ...860L..11K}.  We identify $\sim 20$ stars as belonging to this overdensity, all with $v_z < 0\kms$. We test the IDs of the core Helmi stream stars provided in \cite{2019A&A...625A...5K} against our framework to find their associated network scores. Of the 40 IDs given, 37 pass our parallax error cut.  Of these, 10 have  $S<0.5$ and 27 have high scores of  $S > 0.5$, with 22 having scores $S>0.9$, meaning that the neural network is confident they are accreted. We do not reconstruct the second (smaller) known cluster of the Helmi stream at $v_z > 0\kms$. 

\begin{table*}[t]
\begin{center}
\footnotesize
\renewcommand{\arraystretch}{1.5}
\begin{tabular}{ C{1 cm} | C{1.5 cm} C{1.5 cm}  C{1.5 cm} | C{1.5 cm} C{1.5 cm} C{1.5 cm} |  C{1.5 cm} C{1.5 cm} }
\Xhline{3\arrayrulewidth}
ID& $v_R$~[km/s] & $v_\phi$~[km/s] & $v_z$~[km/s] & $\sigma_R$~[km/s] & $\sigma_\phi$~[km/s] & $\sigma_z$~[km/s] & $N_\text{stars}$  & Frequency \\
 \hline 
I & 21 &  141 & -286 & 40-77 &  8-26 &  8-24  & 10-20 & $100\%$ \\ 
II &  -154 & -377 &  -63 & 63 & 34 & 18 & 3-19 & $95\%$ \\
III &  -195 & -237 &  164 &  41 &  24 & 24  & 3-17 & $78\%$ \\
IV &  212 & -229 &  161 &  9-35 &  7-25 &  1-37  & 7-17 & $74\%$ \\
\Xhline{3\arrayrulewidth}
\end{tabular}
\caption{Centers of the velocity clusters identified by \texttt{DBSCAN} after 100 iterations scanning over position and velocity errors of the stars.  For each cluster, we provide the following: ID number, mean velocities in Galactocentric cylindrical coordinates, range of dispersions of stars that belong in the group through error resampling, number of stars, and its frequency across 100 reruns. We only list groups  that have occurred more then $70\%$  of the time. The velocities of Group~I overlaps with the Helmi stream~\citep{1999Natur.402...53H}, while Groups II, III, and IV may be related to streams identified in~\cite{2018ApJ...860L..11K}. The same groups appear when analyzing either the high-purity or the  canonical sample. \vspace{7pt}}
\label{tab:small_structures}
\end{center}
\end{table*}

The other  three robust streams are located at  $(v_R, v_\phi, v_z) = (-154, -377, -63)\kms$,  $(v_R, v_\phi, v_z) = (-195, -237, 164)\kms$ and  $ (212, -229, 161)\kms$ with small dispersions,\footnote{For groups  II and III, we only provide the maximum dispersions. In some cases, the group has only two stars, making it difficult to define a meaningful velocity dispersion.} as listed in \Tab{tab:small_structures}.  These structures fall near several velocity clusters identified in~\cite{2018ApJ...860L..11K} and may be related. 

\section{Conclusions}
\label{sec:conclusions}

Stars in the Milky Way galaxy can be divided into two components: those that were born within the Galaxy and those that were accreted.  The phase-space distribution of accreted stars provides a crucial handle for understanding how the Galaxy evolved by revealing the imprints of satellite mergers.  This approach requires distinguishing the population of accreted stars from their \emph{in situ} counterparts, a task that becomes increasingly challenging near the Galactic midplane where disk stars comprise $\sim 99$\% of all stars.  This motivated the work of~\cite{ml_paper}, where we developed a deep neural network based approach that allows us to build a catalog of accreted stars from \Gaia DR2 data.  Although the network provides a catalog of all well-measured \Gaia stars, regardless of whether or not they have a line-of-sight velocity measurement, this paper provides the first analysis of the 4.8 million star subset that includes the full 6D information, and that fall within $r \in [6.5, 9.5]\kpc$ and $|z| < 3\kpc$.

Our primary goal is to identify and analyze structures in 3D velocity space.  As a first step, we perform a Gaussian mixture analysis to break down the high-purity sample ($S>0.95$) into its most significant contributions: \Gaia Enceladus, Nyx, and the Halo.  We find that Enceladus is highly radial and comprises the vast majority of accreted stars in this region of the sky.  These results are consistent with previous studies of Enceladus, which characterized its properties farther from the disk plane \citep{2018MNRAS.478..611B,2018Natur.563...85H, 2018arXiv180704290L,2019ApJ...874....3N,2018ApJ...862L...1D}.  Nyx is a new stream identified by this analysis.  It is prograde and comprises nearly  9\% of the high-purity sample, making it one of the most significant streams to be discovered to date near the Sun.  Properties and a discussion of the potential origin of Nyx as a merging dwarf galaxy are explored in greater detail in~\cite{nyx_paper}.  The ``Halo'' is essentially the remaining group of accreted stars that cannot be further subdivided by the mixture analysis. 

We also repeated the analysis on all accreted stars with network scores  $S>0.85$.  This canonical sample is considerably larger in size than the high-purity one, but likely has more contamination from disk stars.  We again recover Enceladus and Nyx in this sample.  Additionally, we find evidence for another prograde stream, Nyx-2, with roughly the same rotational speed as Nyx and equal --- but opposite --- radial velocity. Nyx-2 comprises  $\sim 22\%$ of the canonical sample,  comparable to the Nyx fraction in this sample.  Similarities in the kinematics and metallicities beetween Nyx and Nyx-2 suggest that they may be related to the same progenitor. 

We also attempt to reconstruct non-Gaussian velocity structures using the \texttt{DBSCAN} algorithm.  We locate four additional streams using this method, one of which coincides with the well-studied Helmi stream \citep{1999Natur.402...53H}.  The Helmi stream is by far the most robust of the  four, consisting of $\sim 20$ stars and identified over all repeated iterations that account for uncertaintities in the stellar velocity measurements. The other  three candidate streams consist of fewer stars, and are recovered  $\sim 70\%$ of the time over repeated iterations of \texttt{DBSCAN}.  These streams are  all retrograde, and may be associated with overdensities identified in~\cite{2018ApJ...860L..11K}.

The analysis presented here demonstrates the power derived from combining advancements in data quality, numerical simulations, and data analysis techniques.  The fact that the catalog reproduces known structures such as \Gaia Enceladus and the Helmi stream validates the utility of this approach.  However, the science case extends much farther than simple validation.  Indeed, the new catalog greatly improves our understanding of the stellar distribution in the ROI studied.  In particular, it clearly demonstrates that Enceladus extends down into the Galactic plane.  It also unearths evidence for a significant new stellar stream (Nyx), a potentially related counterpart (Nyx-2), and  three other smaller candidate streams.  

We have improved over previous approaches as a result of two main factors.  The first is due to the statistical benefit of an increased overall size of the accreted stellar sample. Additionally, having used a deep network that is only trained on phase space allows us to derive a high-purity sample of accreted stars without imposing strong cuts on circular velocity or metallicity, as is typically done.  This reduces the intrinsic bias that results from such cuts.
%
We note that this paper has only scratched the surface, and in particular it will be very interesting to investigate what structures can be identified in the rest of our \Gaia DR2 catalog that does not include line-of-sight velocities. 

Understanding the mergers that contributed stellar debris in our neighborhood of the Milky Way has the potential to provide an empirical determination of the local dark matter distribution, which is also built up from mergers~\citep{Herzog-Arbeitman:2017fte, Necib:2018igl}, and as such, is expected to include remnant structures.  The recent discovery of Enceladus, for example, motivates extending the Standard Halo Model of dark matter to (at least) a two-component model that includes both an isotropic halo and debris flow~\citep{2019ApJ...874....3N}.  By clearly demonstrating that Enceladus extends into the disk plane, the results of this work confirm that dark matter debris from this merger likely contributes in the Solar Neighborhood.  The discovery of Nyx near the Solar position may suggest the presence of a corresponding dark matter stream or disk.  Coupling this catalog with cosmological simulations will be essential in refining our understanding of the local dark matter phase-space distribution, and its implications for direct detection experiments.

\section*{Note Added}

As this work was being completed, the paper by \citet{2019arXiv190702527B} became available. Using \texttt{DBSCAN} applied to the integrals of motion, \citet{2019arXiv190702527B} found the groups of stars we also identify in \Tab{tab:small_structures}, along with four more that do not pass our selection criteria. 

\section*{Acknowledgments}

We thank G. Brova, P. Hopkins, E. Kirby, R. Sanderson, and A. Wetzel for helpful discussions.  This work was performed in part at Aspen Center for Physics, which is supported by National Science Foundation grant PHY-1607611. This research was supported by the Munich Institute for Astro- and Particle Physics (MIAPP) of the DFG cluster of excellence ``Origin and Structure of the Universe."  This research was supported in part by the National Science Foundation under Grant No. NSF PHY-1748958.

LN is supported by the DOE under Award Number
DESC0011632, the Sherman Fairchild fellowship, and the California Presidential fellowship.
BO and TC are supported by the US Deptartment of Energy under grant number DE-SC0011640.
ML is supported by the DOE under Award Number
DESC0007968 and the
Cottrell Scholar Program through the Research Corporation for Science Advancement.
MF is supported by the Zuckerman STEM Leadership Program and in part by the DOE under grant number DE-SC0011640.
SGK is supported by an Alfred P. Sloan Research Fellowship, NSF Collaborative Research Grant \#1715847 and CAREER grant \#1455342, and NASA grants NNX15AT06G, JPL 1589742, 17-ATP17-0214.

This work has made use of data from the European Space Agency (ESA) mission Gaia (\url{http://www.cosmos.esa.int/gaia}), processed by the Gaia Data Processing and Analysis Consortium (DPAC, \url{http://www.cosmos.esa.int/web/gaia/dpac/consortium}). Funding for the DPAC has been provided by national institutions, in particular the institutions participating in the Gaia Multilateral Agreement.

\software{
This analysis made use of \texttt{Astropy} \citep{astropy:2018}, \texttt{Galpy} \citep{2015ApJS..216...29B}, \texttt{Matplotlib} \citep{matplotlib}, \texttt{NumPy} \citep{numpy}, and \texttt{Scikit-Learn} \citep{scikit-learn}.
The neural network used for tagging the accreted stars was implemented in \texttt{Keras}~\citep{chollet2015keras} with the \texttt{TensorFlow} backend \citep{tensorflow2015-whitepaper}. The network was trained using \texttt{Adam} \citep{2014arXiv1412.6980K} to minimize the binary cross entropy loss.
}

\def\bibsection{}
\bibliographystyle{aasjournal}
\bibliography{Gaia_ML_2}


\onecolumngrid

\newpage

\appendix
In this Appendix, we show two dimensional histograms of stars in $v_y$--$\sqrt{v_x^2 + v_z^2}$ space with $S<0.05$ and $S\in [0.3,0.5]$ in \Fig{fig:disk_hists}.  We find that the scores roughly track the different components of the sample, with $S < 0.05$ largely identifying the thin disk, and $S \in [0.3, 0.5]$ largely identifying the thick disk. Thick-disk stars are harder to discern from the thin disk and the (largely accreted) halo and thus receive mid-range scores.  Remember that the neutral network is only identifying \emph{in-situ} and accreted stars---the fact that the network scores roughly correlate with physical components of the Galaxy is an added bonus. In \Fig{fig:ratios_corner} and \Fig{fig:ratio_canonical}, we provide the corner plots of the fractions of each component in the Gaussian mixture model analysis for the high-purity and canonical samples, respectively. In \Fig{fig:nyx_corner} (\ref{fig:nyx_corner_75}), we show the best-fit parameters of Nyx in the high-purity (canonical) sample. The best-fit parameters of Nyx-2 in the canonical sample are shown in \Fig{fig:nyx2_corner_75}. Similar treatment of Enceladus is shown in \Figs{fig:subs_corner}{fig:subs_corner_75} for the high-purity and canonical sample,  respectively. The residual of the Gaussian mixture analysis of the high-purity sample is shown in \Fig{fig:residuals}.  In \Fig{fig:sausage_spatial}, we provide the spatial distribution of Enceladus stars, and focus particularly on how they extend down to the Galactic midplane. In Figs.~\ref{fig:orbit_nyx}--\ref{fig:orbit_thick}, we provide the orbital properties of Nyx, Enceladus, and Halo for the high-purity sample, as well as stars with scores $S<0.05$ and $S\in[0.3, 0.5]$ for reference.

\setcounter{equation}{0}
\setcounter{figure}{0}
\setcounter{table}{0}
\setcounter{section}{0}
\makeatletter
\renewcommand{\theequation}{S\arabic{equation}}
\renewcommand{\thefigure}{S\arabic{figure}}
\renewcommand{\thetable}{S\arabic{table}}

\begin{figure*}[h]
\centering
\includegraphics[width=0.95\textwidth]{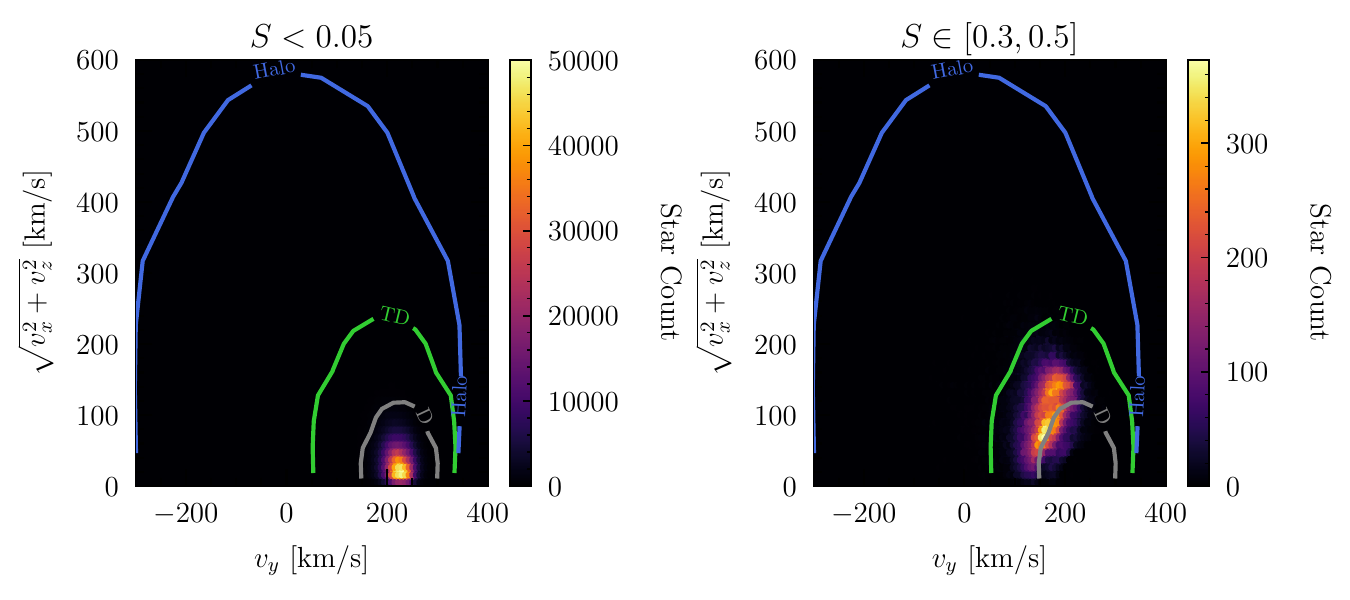}
\caption{Toomre plots for stars with scores $S< 0.05$ (left) and $S \in [0.3, 0.5]$ (right) along with the $3\sigma$ velocity contours for the thin disk (D), thick disk (TD), and stellar halo (Halo), following \cite{2003A&A...410..527B}. }
\label{fig:disk_hists}
\end{figure*}

\setcounter{figure}{0}
\makeatletter
\renewcommand{\thefigure}{S2\alph{figure}}

\begin{figure*}[h]
\centering
\includegraphics[width=0.45\textwidth]{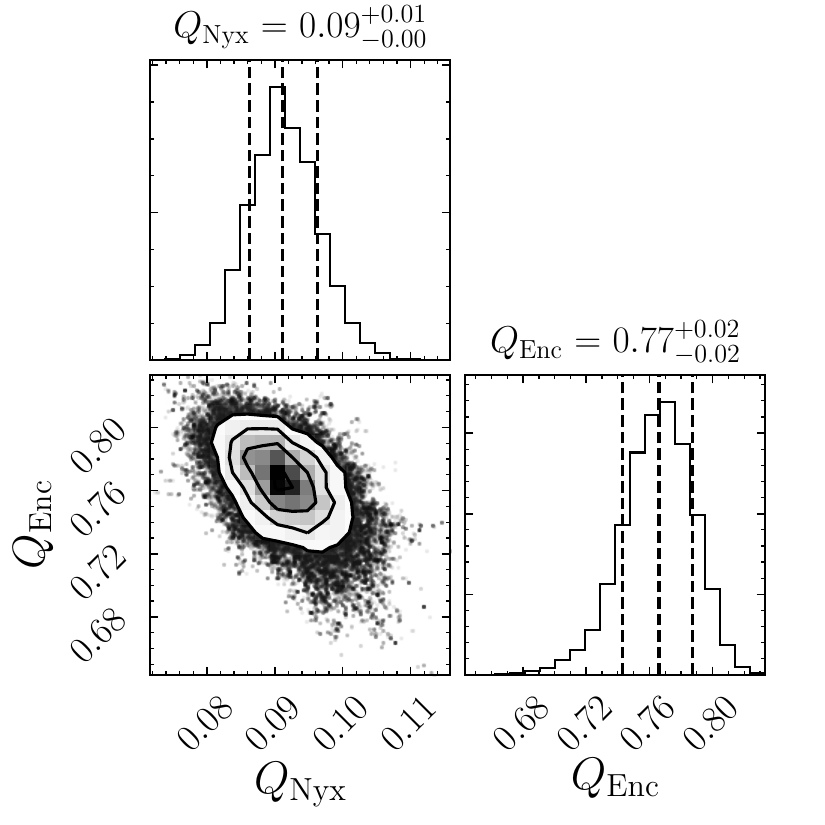}
\caption{Corner plot of the \texttt{emcee} run for the fractions of the different components in the high-purity sample.  }
\label{fig:ratios_corner}
\end{figure*}

\begin{figure*}[h]
\centering
\includegraphics[width=0.65\textwidth]{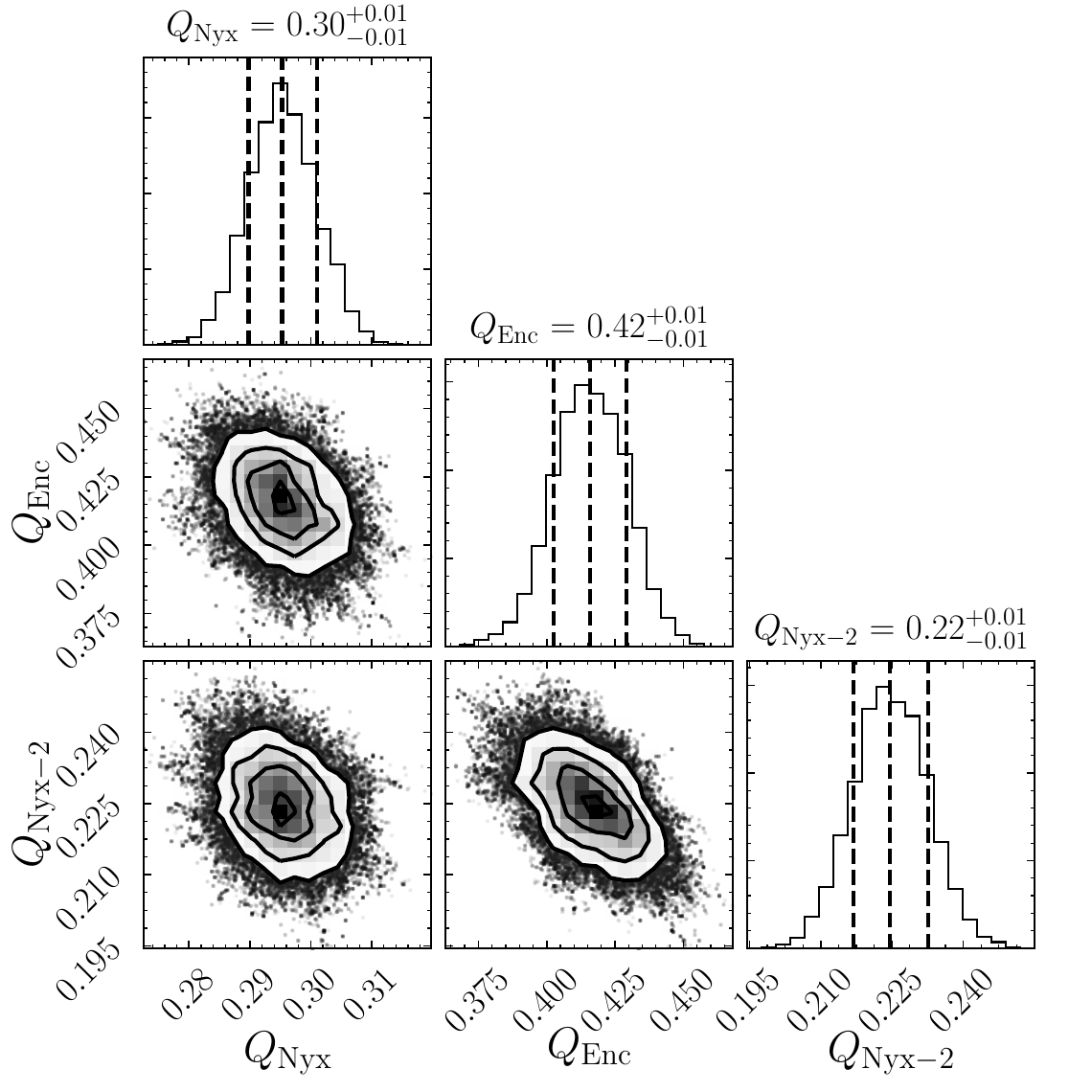}
\caption{Corner plot of the \texttt{emcee} run for the fractions of the different components in the canonical sample.  }
\label{fig:ratio_canonical}
\end{figure*}

\begin{figure*}[h]
\centering
\includegraphics[width=0.95\textwidth]{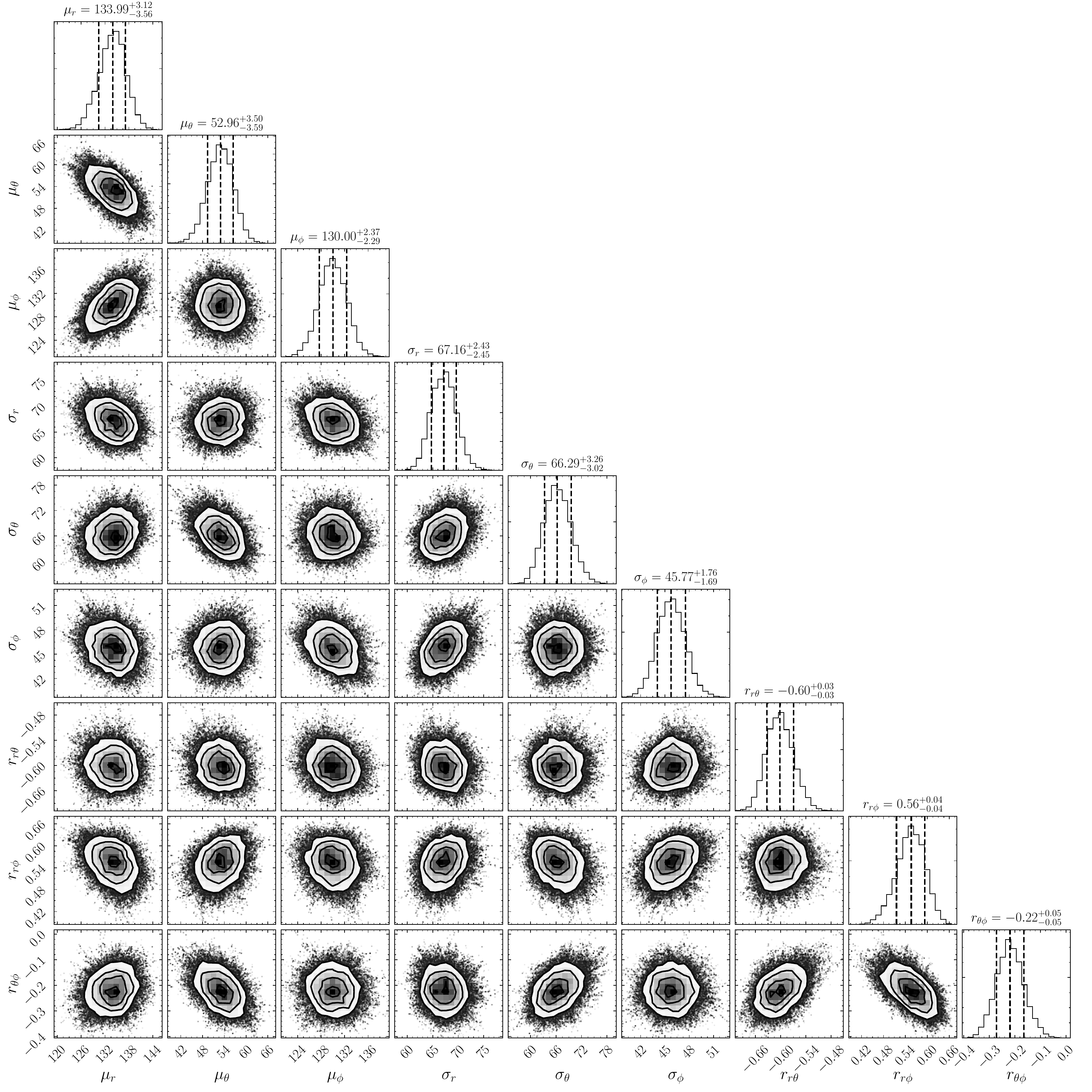}
\caption{Corner plot of the \texttt{emcee} run for Nyx in the high-purity sample.}
\label{fig:nyx_corner}
\end{figure*}

\begin{figure*}[h]
\centering
\includegraphics[width=0.95\textwidth]{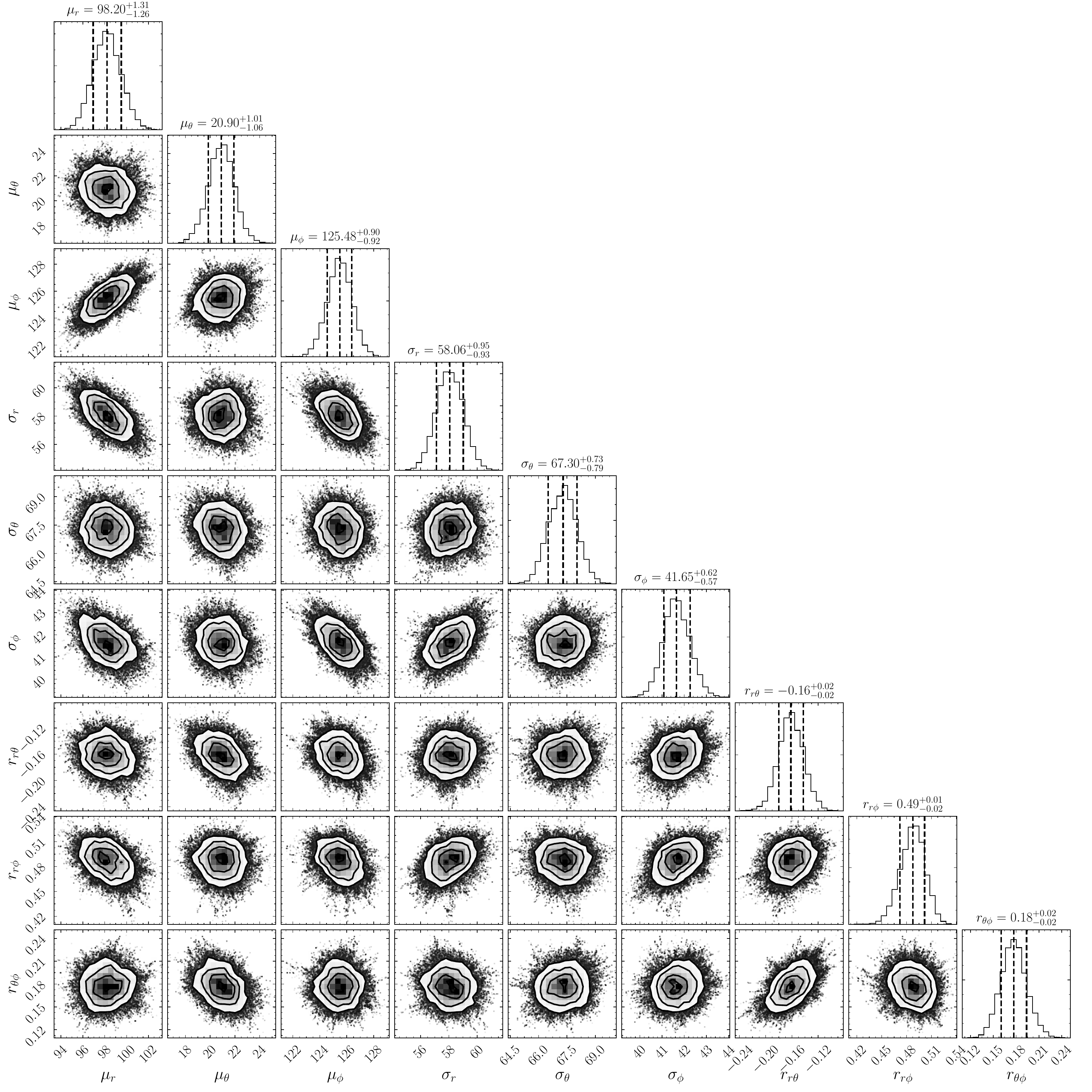}
\caption{Corner plot of the \texttt{emcee} run for Nyx in the canonical sample. }
\label{fig:nyx_corner_75}
\end{figure*}

\begin{figure*}[h]
\centering
\includegraphics[width=0.95\textwidth]{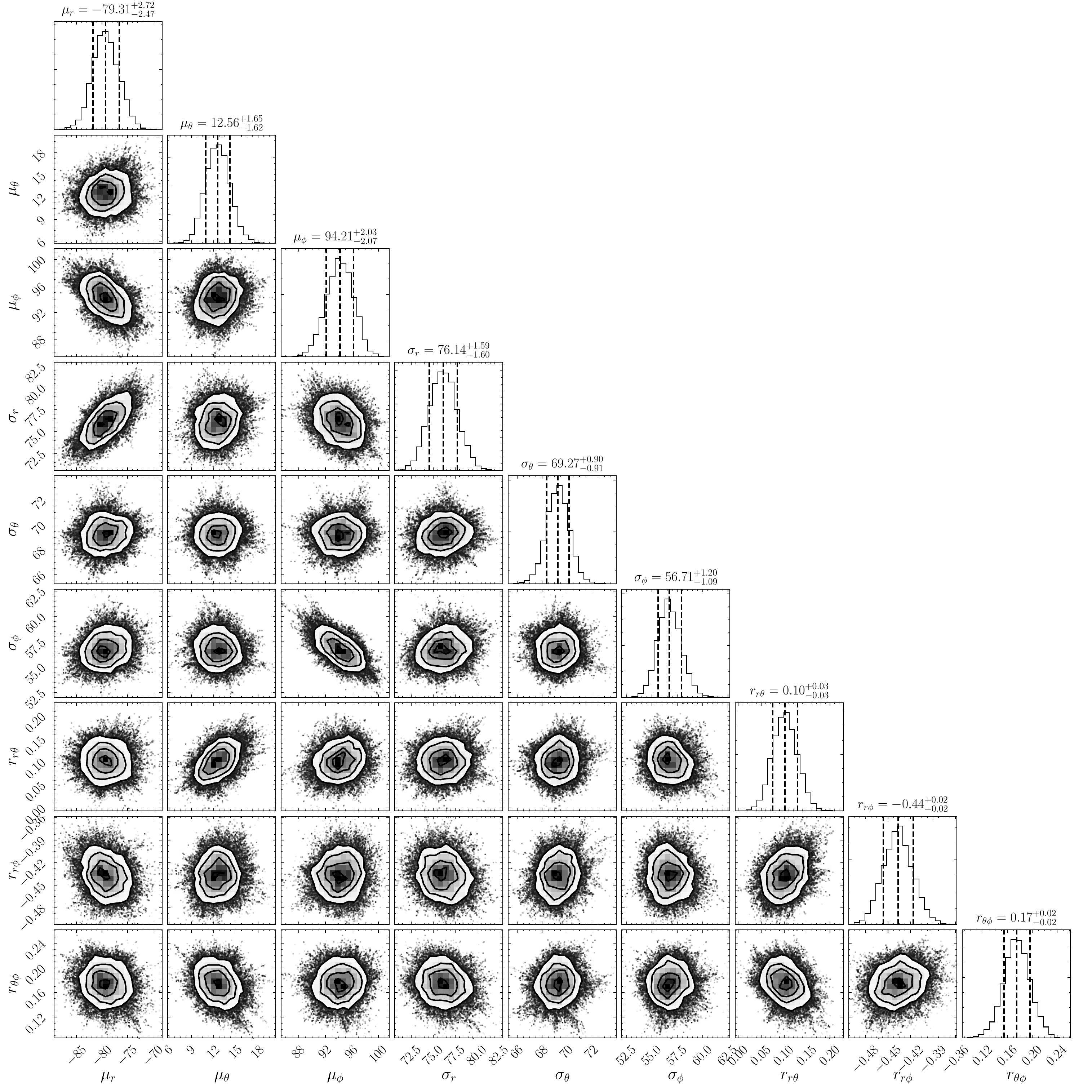}
\caption{Corner plot of the \texttt{emcee} run for Nyx-2 in the canonical sample. }
\label{fig:nyx2_corner_75}
\end{figure*}

\begin{figure*}[t]
\centering
\includegraphics[width=0.95\textwidth]{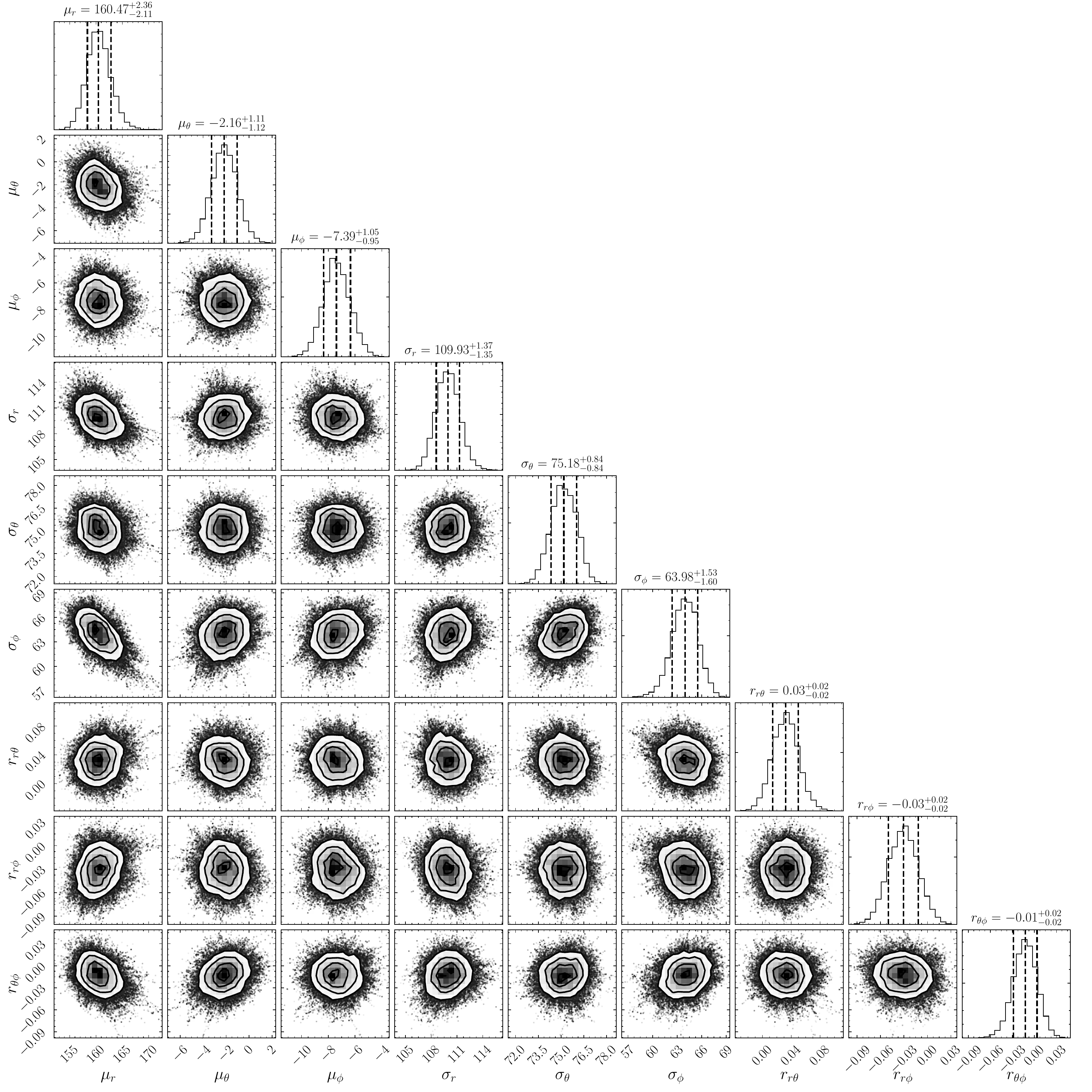}
\caption{Corner plot of the \texttt{emcee} run for Enceladus in the high-purity sample. }
\label{fig:subs_corner}
\end{figure*}

\begin{figure*}[t]
\centering
\includegraphics[width=0.95\textwidth]{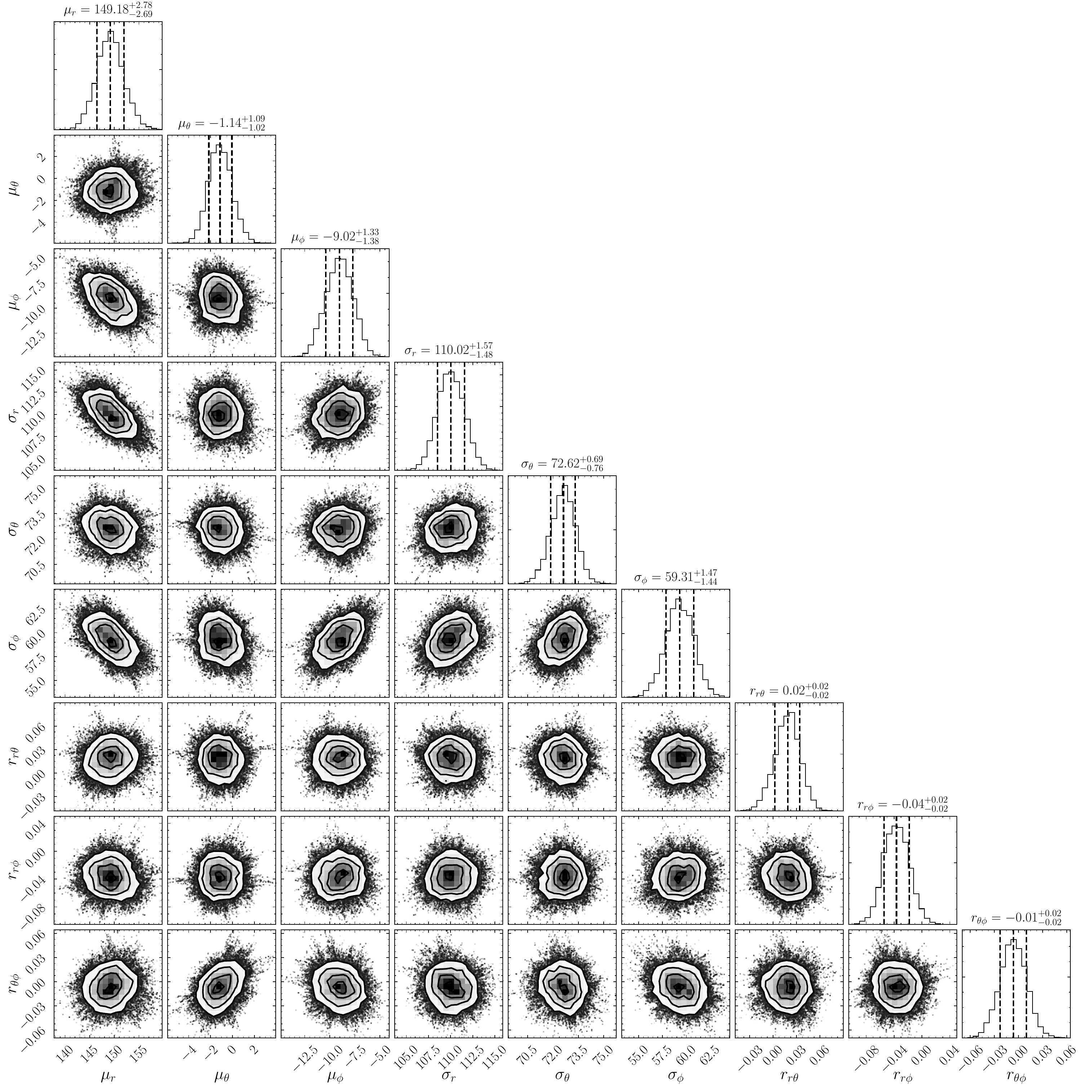}
\caption{Corner plot of the \texttt{emcee} run for Enceladus in the canonical sample. }
\label{fig:subs_corner_75}
\end{figure*}

\setcounter{figure}{2}
\makeatletter
\renewcommand{\thefigure}{S\arabic{figure}}

\begin{figure}[t]
\centering
\includegraphics[width=0.8\textwidth]{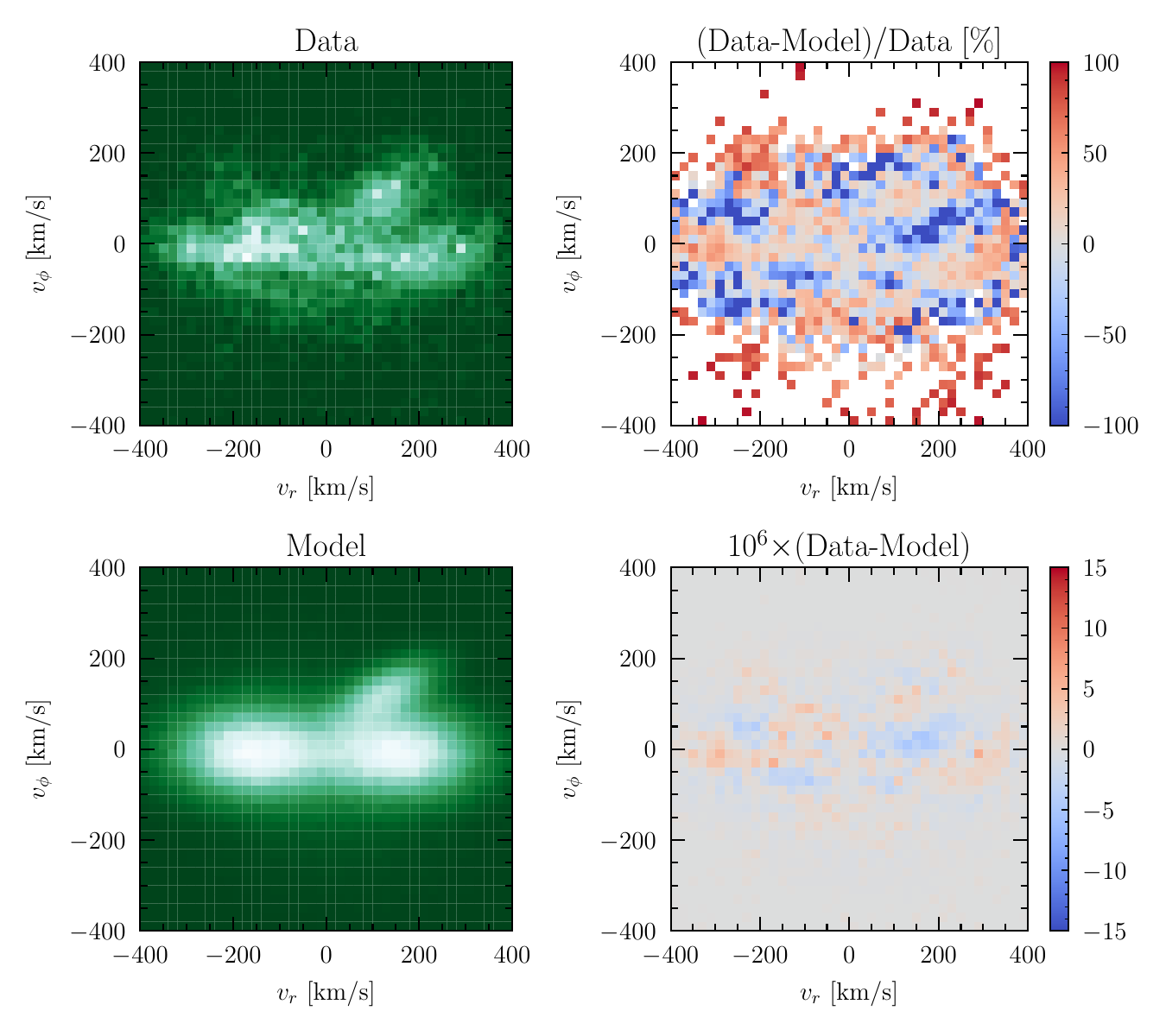}
\caption{Residuals for the high-purity sample.  The top left panel shows the data in the $v_r - v_\phi$ plane.  The bottom left panel shows the model prediction.  The right panels show the residuals ($\text{Data}-\text{Model}$); these are normalized to the data count in the top panel.  }
\label{fig:residuals}
\end{figure}

\begin{figure*}[t]
\centering
\includegraphics[width=0.95\textwidth]{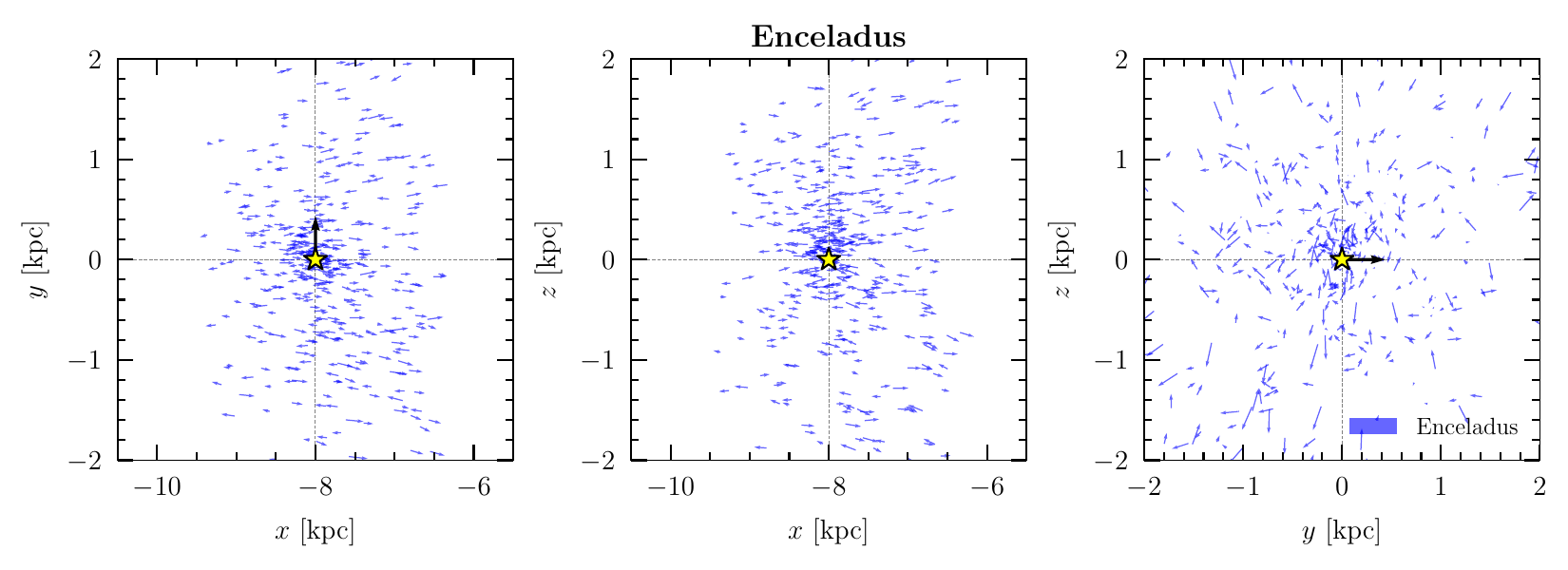}
\caption{  
Spatial distribution of the velocity vectors for the  stars associated with \Gaia Enceladus in the high-purity sample. The Sun is located at $(x,y,z) = (-8, 0,0)\kpc$, and the black arrow indicates its velocity. The sample of Enceladus stars has been subsampled by a factor of 5 for ease of viewing. }
\label{fig:sausage_spatial}
\end{figure*}

\setcounter{figure}{0}
\makeatletter
\renewcommand{\thefigure}{S5\alph{figure}}

\begin{figure*}[t] 
\centering
\includegraphics[width=0.95\textwidth]{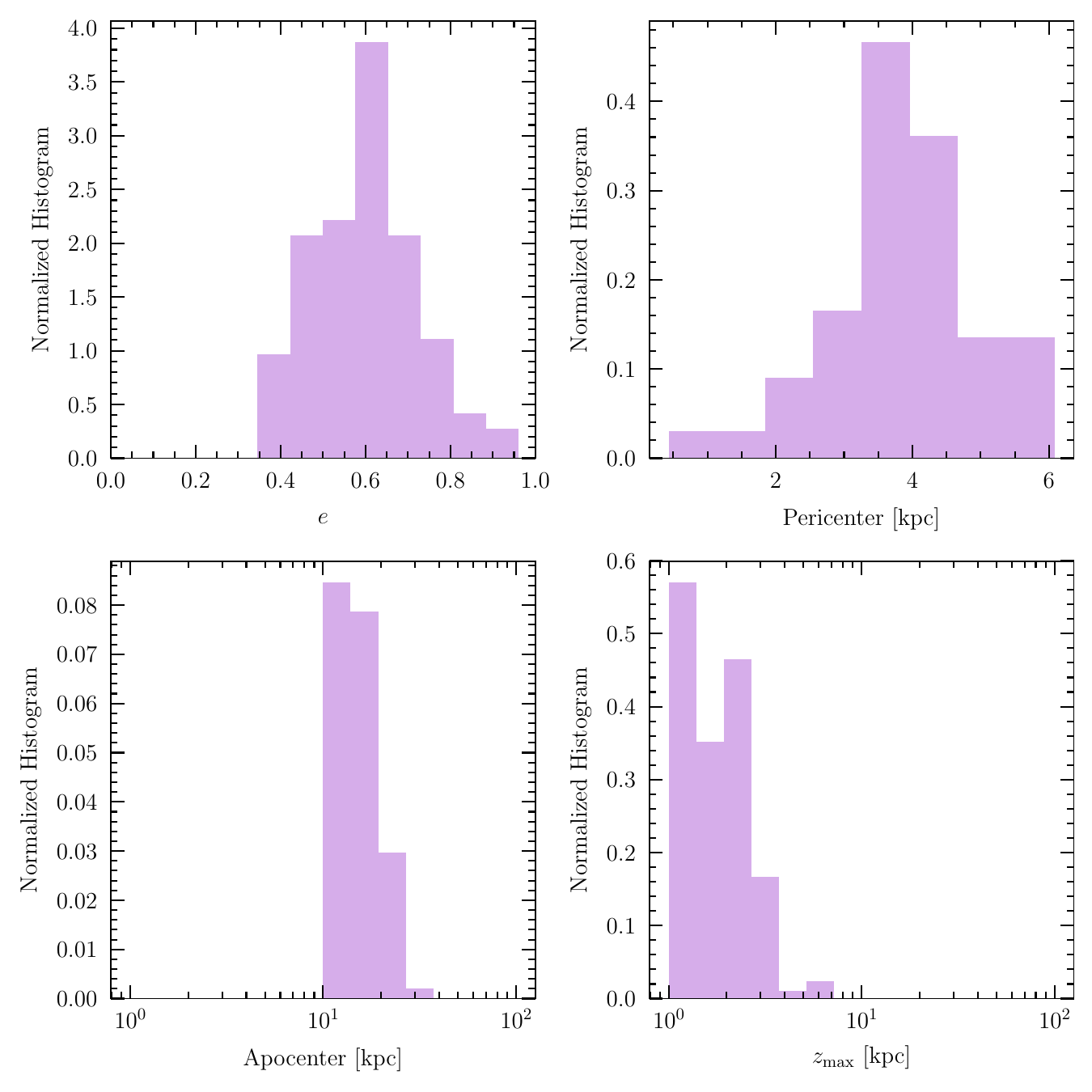}
\caption{  Orbital properties of stars associated with Nyx, calculated as described in the main text.  We show distributions for the eccentricities (top left), pericenters (top right), apocenters (bottom left), and $z_{\rm{max}}$ (bottom right). The orbits were evolved back 1 Gyr over 1000 steps using \texttt{gala} \citep{gala}, assuming the default Milky Way potential in \cite{2015ApJS..216...29B}. }
\label{fig:orbit_nyx}
\end{figure*}

\begin{figure*}[t]
\centering
\includegraphics[width=0.95\textwidth]{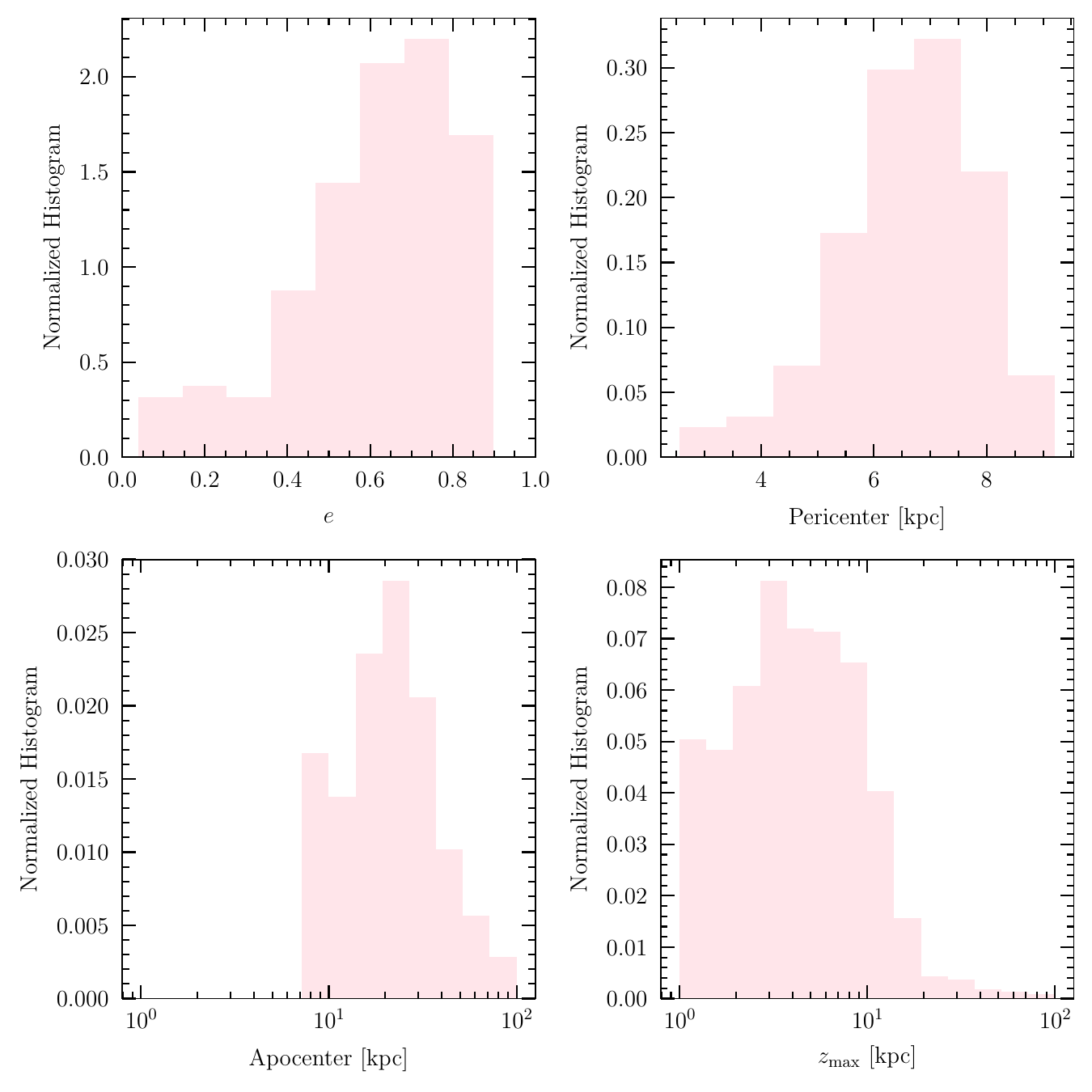}
\caption{Same as \Fig{fig:orbit_nyx}, except for the halo stars.  }
\label{fig:orbit_halo}
\end{figure*}

\begin{figure*}[t]
\centering
\includegraphics[width=0.95\textwidth]{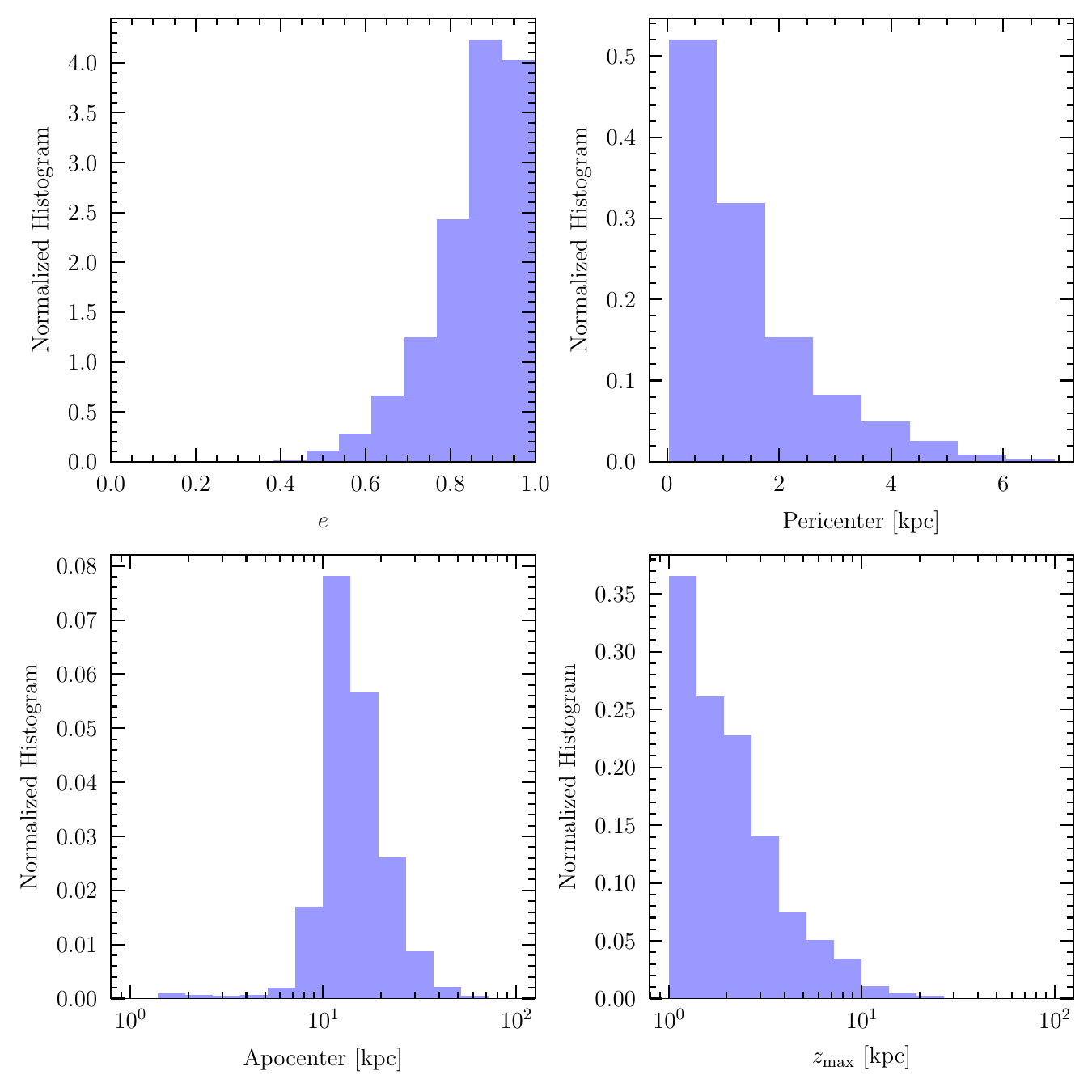}
\caption{Same as \Fig{fig:orbit_nyx}, except for the Enceladus stars.  }
\label{fig:orbit_sausage}
\end{figure*}

\begin{figure*}[t]
\centering
\includegraphics[width=0.95\textwidth]{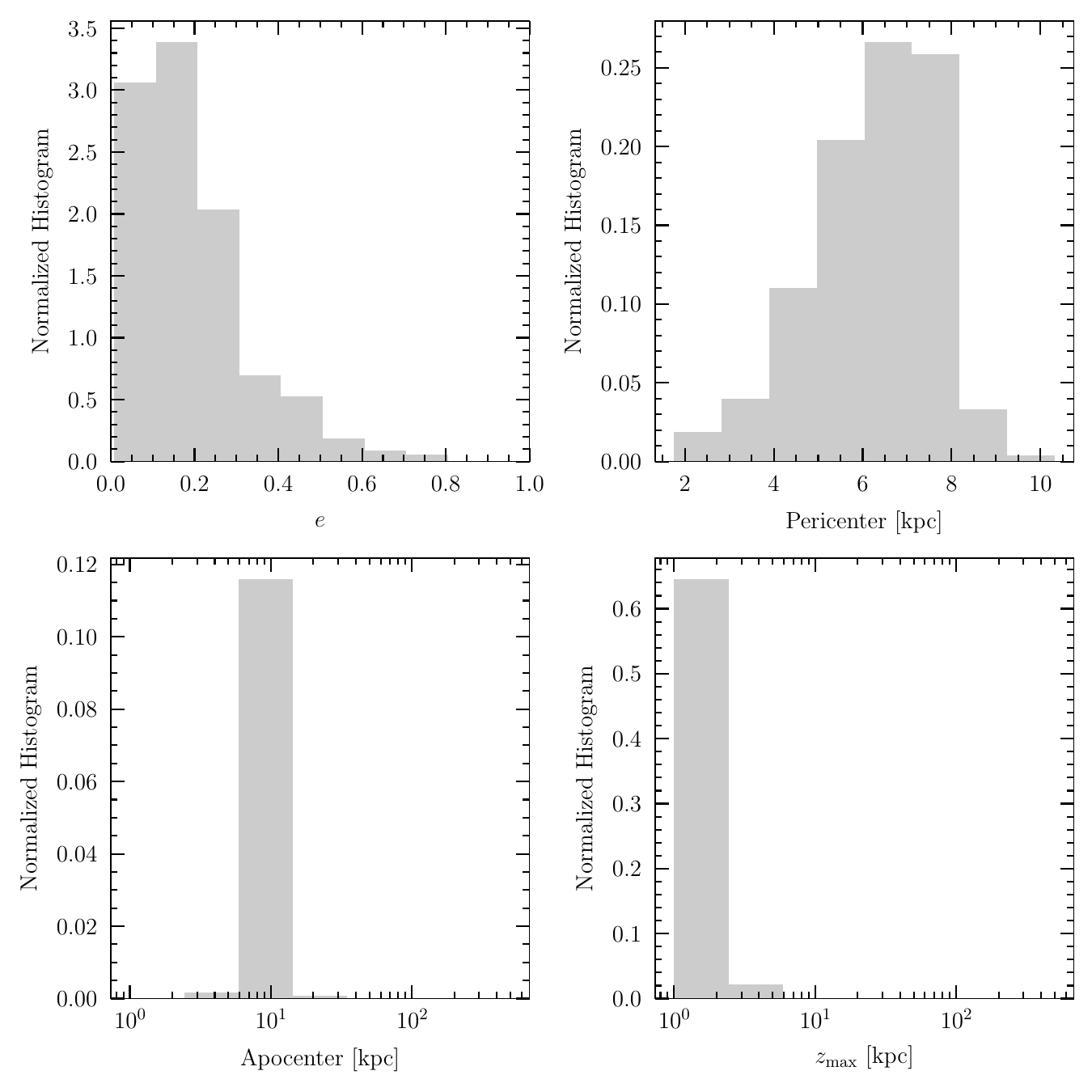}
\caption{Same as \Fig{fig:orbit_nyx}, except for stars with network scores $S<0.05$. }
\label{fig:orbit_thin}
\end{figure*}

\begin{figure*}[t]
\centering
\includegraphics[width=0.95\textwidth]{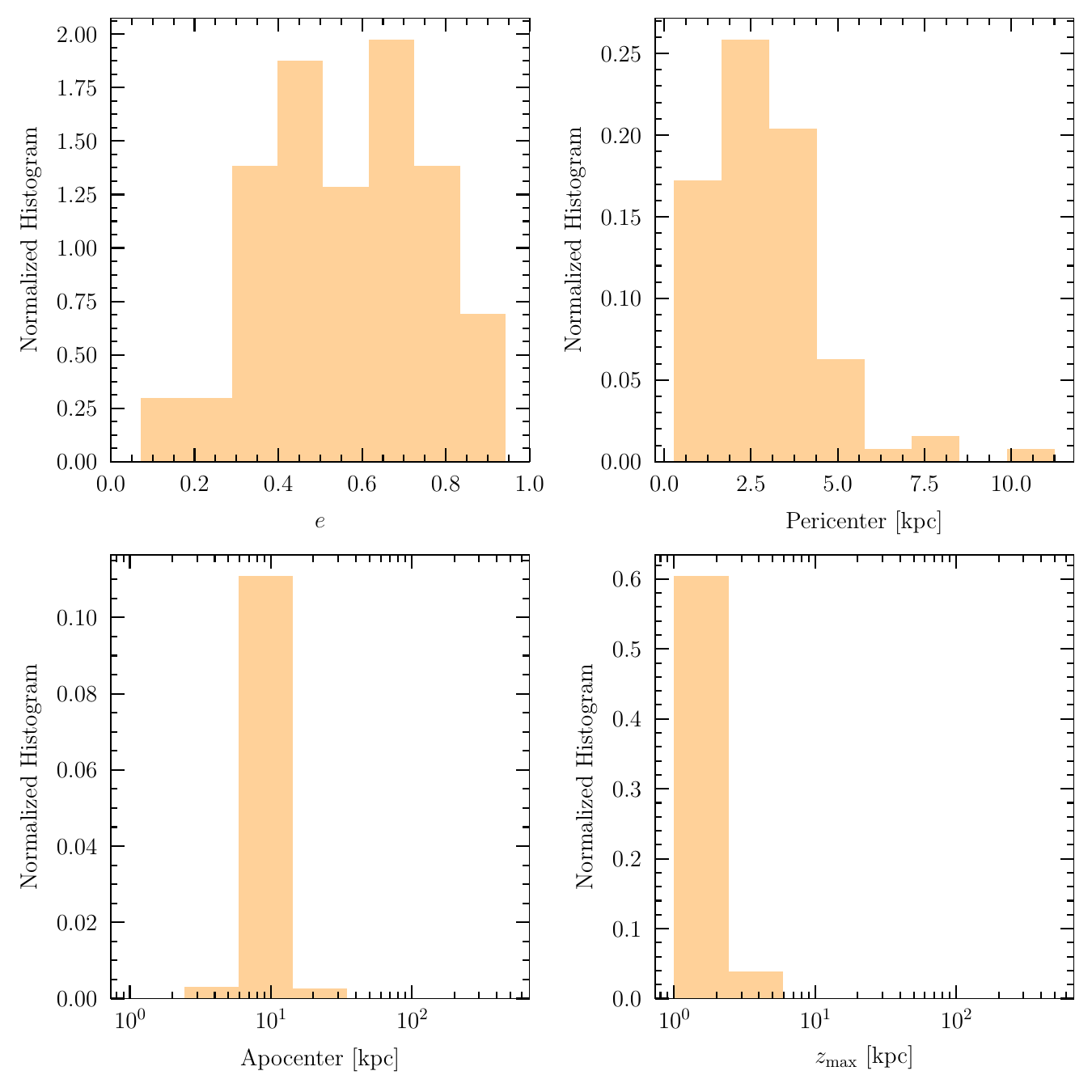}
\caption{Same as \Fig{fig:orbit_nyx}, except for stars with network scores $S \in [0.3, 0.5]$.  }
\label{fig:orbit_thick}
\end{figure*}

\end{document}